\documentclass[twocolumn,tighten,dvipsnames,twocolappendix]{aastex701}

\usepackage{amsmath,amssymb,mathtools,mathrsfs}

\usepackage{xspace}

\def\muas{\mu{\rm as}}

\newcommand{\C}{\texttt{C}\xspace}

\definecolor{ForestGreen}{rgb}{0.133,0.545,0.133}

\definecolor{DarkBlue}{rgb}
{0.0,0.0,0.545}

\newcommand{\erg}{{\rm erg}}
\newcommand{\s}{{\rm s}}

\newcommand{\K}{{\rm K}}

\def\Mpc{{\rm Mpc}}

\def\Fopt{F_{\rm opt}}
\def\Fbol{F_{\rm bol}}
\def\Lbol{L_{\rm bol}}
\def\Lpho{L_{\rm pho}}
\def\eacc{\eta_{\rm acc}}
\def\eadv{\eta_{\rm adv}}
\def\MBH{M}
\def\mAA{{\rm\AA}}

\defcitealias{M87PaperI}{M87*~Paper~I}
\defcitealias{M87PaperII}{M87*~Paper~II}
\defcitealias{M87PaperIII}{M87*~Paper~III}
\defcitealias{M87PaperIV}{M87*~Paper~IV}
\defcitealias{M87PaperV}{M87*~Paper~V}
\defcitealias{M87PaperVI}{M87*~Paper~VI}
\defcitealias{M87PaperVII}{M87*~Paper~VII}
\defcitealias{M87PaperVIII}{M87*~Paper~VIII}
\defcitealias{M87PaperIX}{M87*~Paper~IX}

\defcitealias{SgrAPaperI}{Sgr~A*~Paper~I}
\defcitealias{SgrAPaperII}{Sgr~A*~Paper~II}
\defcitealias{SgrAPaperIII}{Sgr~A*~Paper~III}
\defcitealias{SgrAPaperIV}{Sgr~A*~Paper~IV}
\defcitealias{SgrAPaperV}{Sgr~A*~Paper~V}
\defcitealias{SgrAPaperVI}{Sgr~A*~Paper~VI}
\defcitealias{SgrAPaperVII}{Sgr~A*~Paper~VII}
\defcitealias{SgrAPaperVIII}{Sgr~A*~Paper~VIII}

\shorttitle{Observational Constraints on Surface Emission in AGNs}
\shortauthors{E.~Oran, A.E.~Broderick \& K.~Salehi}

\begin{document}

\title{Observational Constraints on Horizonless Compact Objects from Thermal Emission in AGNs}

\author[0009-0000-3204-8808]{Ekin Oran}
\email{eoran@uwaterloo.ca}
\affiliation{Department of Physics and Astronomy, University of Waterloo, 200 University Avenue West, Waterloo, ON, N2L 3G1, Canada}

\author[0000-0002-3351-760X]{Avery E. Broderick}
\email{abroderick@perimeterinstitute.ca}
\affiliation{Department of Physics and Astronomy, University of Waterloo, 200 University Avenue West, Waterloo, ON, N2L 3G1, Canada}
\affiliation{Perimeter Institute for Theoretical Physics, 31 Caroline Street North, Waterloo, ON, N2L 2Y5, Canada}
\affiliation{Waterloo Centre for Astrophysics, University of Waterloo, Waterloo, ON N2L 3G1 Canada}

\author[0009-0006-0070-1888]{Kiana Salehi}
\email{k2salehi@uwaterloo.ca}
\affiliation{Department of Physics and Astronomy, University of Waterloo, 200 University Avenue West, Waterloo, ON, N2L 3G1, Canada}
\affiliation{Perimeter Institute for Theoretical Physics, 31 Caroline Street North, Waterloo, ON, N2L 2Y5, Canada}
\affiliation{Waterloo Centre for Astrophysics, University of Waterloo, Waterloo, ON N2L 3G1 Canada}

\begin{abstract}
The Swift Burst Alert Telescope Active Galactic Nuclei Spectroscopic Survey (BASS DR2) provides one of the largest and most complete samples of bright, local ($z < 0.1$) active galactic nuclei (AGNs) with high-quality spectroscopic measurements. These sources are typically interpreted within the framework of Kerr black holes; however, a wide range of horizonless compact-object scenarios—including naked singularities, black hole mimickers, and other exotic compact objects—have been proposed as alternatives.
Largely independent of the physics of the particular horizonless model invoked, an accretion powered, thermally radiating photosphere is expected to develop on astronomically short timescales.  
In this paper, we investigate whether the presence of such a photosphere is consistent with the observed properties of an appropriate subsample of BASS AGNs. We find that in no instances is such a spectral feature present, and can exclude its existence in significant fraction of objects.  For the remaining sources, modest improvements in observational sensitivity and spectral coverage could enable robust exclusion. Our results extend previous horizon-scale tests performed for M87* and Sgr A* to a large AGN population, providing strong, population-level evidence against horizonless alternatives.  We conclude by outlining future observational strategies that can further tighten these constraints and significantly increase the number of objects for which horizons are required.

\end{abstract}


\section{Introduction}

Active galactic nuclei (AGNs) provide some of the most luminous and well-studied environments in the universe, offering a unique laboratory for probing the nature of compact objects in strong-gravity regimes. The Swift Burst Alert Telescope (BAT) Active Galactic Nuclei Spectroscopic Survey (BASS) provides a comprehensive spectroscopic characterization of hard-X-ray-selected AGNs in the local universe \citep{Koss2022b}. The second data (DR2) release of the survey includes 1449 optical spectra for 858 AGNs drawn from the Swift BAT 70-month all-sky sample, complemented by extensive optical and near-infrared spectroscopic follow-up. BASS further delivers a uniformly processed master catalog containing redshifts, distances, bolometric luminosities, and compact object mass estimates, enabling detailed population-level studies \citep{Koss2022}. Owing to its high completeness and well-characterized selection, BASS provides a robust observational foundation for probing the nature of compact objects powering local AGNs.

Within the framework of general relativity (GR), the compact objects powering active galactic nuclei are conventionally modeled as supermassive black holes (SMBHs) with masses in the range $\sim 10^{7}$--$10^{9}\,M_\odot$. This paradigm is supported by a broad body of theoretical and observational work dating back to early studies of quasars and galactic nuclei
\citep[e.g.,][]{LyndenBell1969,LyndenBell1971,Rees1984, M87_PaperI}, which demonstrated that the large energy output of AGNs can be naturally explained by the release of gravitational binding energy through accretion onto a compact object. Additional evidence comes from population-level arguments, such as the argument mentioned in \citet{Soltan1982}, which uses the integrated quasar luminosity density to infer the total mass accumulated in black holes, showing that the observed energy output of quasars implies a substantial cosmological density of SMBHs with masses $\sim 10^{8}\,M_\odot$. This result is largely independent of the underlying cosmological model and provides strong evidence that luminous AGNs are powered by accretion onto massive compact objects. Additionally, recent horizon-scale imaging by the Event Horizon Telescope provides further support for this picture \citep{M87_PaperI,M87_PaperII,M87_PaperIII,M87_PaperIV,M87_PaperV,M87_PaperVI,SgpaperI,SgpaperII,SGpaperIII,SgpaperIV,SgpaperV,SgpaperVI}.

Despite this compelling evidence, most observational probes of AGNs test the gravitational potential rather than the existence of an event horizon itself. A defining feature of a black hole is the existence of an event horizon, which hides a central spacetime singularity from external observers. While the singularity itself signals the breakdown of classical GR and likely requires a quantum theory of gravity for its complete description, the horizon ensures that this pathology remains observationally inaccessible. However, GR also admits solutions in which singularities are not hidden behind horizons, giving rise to naked singularities, for example in extremal or over-extremal configurations with sufficiently large spin or charge. More broadly, a wide class of horizonless compact-object models—including black hole mimickers such as boson stars, fermion stars, and gravastars—has been proposed, many of which can reproduce black-hole--like gravitational potentials. 

A key and robust consequence of the absence of an event horizon\footnote{The teleological nature of the event horizon makes it difficult to constrain observationally.  Thus, hereafter we will use this interchangeably with the physically less salient but practically more useful notion of an astrophysical horizon, which may encompass both apparent horizons and objects in which material is completely hidden for cosmologically long timescales.} is the formation of a radiating photosphere powered by accretion which produces a thermal emission component that is largely independent\footnote{Some authors have explored ways in which thermalization may be avoided \citep[e.g.][]{Cardoso2019,Carballo2023}, though the physics of the accreting baryonic gas provides a natural way to avoid those complications in many instances \citep{Faraji2026}.} of the specific underlying model \citep{NarayanHeyl2002,Broderick2006,Broderick2009,Broderick2015,Broderick2024,Faraji2026}. This provides a model-agnostic observational signature that can be used to test the existence of horizons. Previous work has applied this framework to horizon-scale sources such as M87$^\ast$ and Sgr A$^\ast$, placing strong constraints on the presence of such photospheres \citep{Broderick2006,Broderick2009,Broderick2015,Broderick2024, M87_PaperVI,SgpaperVI}.

In this work, we extend this approach to a large population of AGNs using the BASS DR2 sample, enabling a systematic, population-level test of horizonless compact-object scenarios. Specifically, we test the existence of an event horizon in the compact objects powering AGNs by searching for excess electromagnetic emission expected in horizonless models. In the absence of an event horizon, energy advected inward by the accretion flow cannot be permanently absorbed and must ultimately be thermalized and re-emitted, producing an observable contribution to the spectral energy distribution (SED) \citep[see, e.g.,][]{Broderick2024,Uniyal2025,Faraji2026}. Conversely, if an event horizon is present, this energy is lost into the black hole and no such emission is expected. By comparing observational constraints with the predicted emission under horizonless scenarios, we assess whether the AGN population is consistent with the existence of an event horizon.

The BASS catalog provides the essential observational inputs required for this test, including reliable redshift measurements and compact object mass estimates, which we use to compute the expected photospheric flux under horizonless scenarios. We model the accretion flow as a radiatively inefficient accretion flow (RIAF), as theoretically anticipated for systems accreting at rates well below the Eddington limit, $\dot{M} \lesssim 0.01\,\dot{M}_{\rm Edd}$.

The resulting emission produces a thermal-like feature in the SED, whose luminosity and characteristic temperature are determined by energy balance with the accretion rate which can be inferred with the information included in the BASS catalog. Detecting or constraining such emission therefore provides a direct observational test of the presence of an event horizon. This paper is organized as follows. In \autoref{sec:2}, we review the general properties of accretion powered photospheres and when and how they appear in accreting spherically symmetric horizonless objects.
In \autoref{sec:3}, we describe the methodology used to test for the presence of an event horizon in AGNs drawn from the BASS catalog. \autoref{sec:4} presents our results and discusses the subset of AGNs that may be consistent with horizonless compact objects. Finally, we summarize our conclusions in \autoref{sec:C}. 

\section{Generic Static, Spherically Symmetric Horizonless Compact Objects}
\label{sec:2}

In general relativity, a black hole is defined by the existence of an event horizon. This is the causal boundary beyond which matter and radiation cannot escape to infinity. Thus, energy that is advected inward by accretion is permanently removed from the observable Universe. This leads to characteristic observational signatures, including a suppression of emission from the innermost regions. In contrast, horizonless compact objects lack an event horizon, allowing matter and radiation to remain, at least in principle, observationally accessible.

Under minimal assumptions of locality and standard microphysical behavior
outside some microscopic distance from the object,
the absence of an event horizon qualitatively changes the behavior of trajectories.
In particular, trajectories that would otherwise be irreversibly captured by a black hole's event horizon may instead either remain bound near the compact object or escape from it. Both possibilities can give rise to distinctive observational consequences.

The fate of matter that remains bound to the central object depends on the putative interactions between the object and the accreting baryonic material. If the surface absorbs the baryons, it necessarily heats and for reasonable accretion histories for known black holes will reach thermodynamic equilibrium on astronomically relevant timescales \citep{Broderick2007B,Broderick2009,Broderick2024}. 

If the surface does not itself heat, the development of a baryonic atmosphere results in an optically thick photosphere \citep{Olivares2020,Faraji2026}.  For objects in which the ingoing trajectories are unbound, for a broad class of horizonless compact object and naked singularity models, the accreting matter can encounter turning points where it reverses direction and interacts with subsequent infalling material, giving rise to shocks \citep{Vieira2023,Broderick2024,Kluzniak2024,Uniyal2025}. In either case, these processes lead to the buildup of a dense matter configuration in the vicinity of the compact object.

The formation of an effective accretion-powered photosphere follows from the dissipation and redistribution of energy within this dense region. Kinetic energy is converted into heat through shocks, compression, and interactions in the flow, and is redistributed via particle collisions, radiation processes, and turbulent mixing, driving the system toward local thermodynamic equilibrium. Depending on the spacetime and accretion dynamics, the emitting region can correspond to a physical surface, a baryon-loaded atmosphere, or a quasi-steady settling flow. In all cases, thermalization and photon diffusion occur on timescales that are short compared to relevant observational timescales unless the spacetime is finely tuned such that the photosphere is located within the small interior region where extreme redshifts develop
\citep[though see][where limits on the practical redshift are discussed in the context of baryonic atmospheres about general black hole mimickers]{Faraji2026}. 

Thus, for a broad class of horizonless compact objects, the accumulated matter is well described by an effective photosphere emitting thermally \citep[see][for further discussion]{NarayanHeyl2002,Broderick2024,Faraji2026}.

We now turn to the properties of this photosphere and its relationship to the observable properties of the accretion flow in AGNs.

\subsection{Photosphere Luminosity}
AGNs are powered by accretion.\footnote{Jet models that tap the spin of the central object necessarily require periodic accretion to inject additional angular momentum.}  The gravitational potential energy of infalling matter is converted into kinetic energy and then radiated, producing broadband emission from the radio through the gamma-rays.  It is natural, then, to express the bolometric luminosity of the accretion flow as a fraction of the mass accretion rate, i.e.,
\begin{equation}
    \Lbol = \eacc \dot{M} c^2,
    \label{eq:Lbol}
\end{equation}
where $\dot{M}$ is the mass accretion rate onto the central object and $\eacc$ is the denotes the fraction of the accreted rest-mass energy liberated and subsequently radiated during infall.  For quasars, typical values of $\eacc$ are $\sim10\%$ \citep{Page1974,Soltan1982}. However, this is predicated on the efficient radiation of the energy accrued by the infalling matter during the accretion process.

At low accretion rates, i.e., $\dot{M}$ less than 1\% of the Eddington limit, accretion proceeds via a radiatively inefficient mode \citep{Narayan1995a,Narayan1995b,Narayan1998,Yuan2014}.  These radiatively inefficient accretion flows (RIAFs) cannot cool, and are therefore typically nearly virial, with temperatures near the central object approaching $10^{11}~\K$, and therefore expected to be nearly completely ionized, typically geometrically thick, and optically thin.  The low densities within the flow due to the low $\dot{M}$ result in weak coupling between the nuclei (which gain most of the energy during infall) and the leptons (more generally, electrons) which can rapidly radiate via synchrotron (radio) and synchrotron self-Compton or Bremsstrahlung (X-rays and gamma-rays).  Thus, in this accretion state, $\eacc$ is small, typically $\lesssim 1\%$.

Because RIAFs cannot efficiently lose energy, they necessarily advect it inwards (hence the advection dominated accretion flow, or ADAF, models).  For black holes, by the time the accreting material reaches the event horizon, nearly 100\% of the rest-mass energy can be converted into kinetic energy.  As described above, should an event horizon be absent, the accreting material will typically collect and produce an optically thick, accretion powered, thermally emitting photosphere. 

For the reasons described above, this photosphere is expected to rapidly reach steady state, radiating the power supplied by accretion.  Therefore, as seen by distant observers the photosphere luminosity is given by
\begin{equation}
    \Lpho = \eadv \dot{M} c^2,    
    \label{eq:Lpho}
\end{equation}
where we conservatively take $\eadv\sim0.5$.

\subsection{Photosphere Temperature and Spectrum}
The properties of a putative photosphere translate into strong predictions for its observational characteristics.  Because it is thermalized and optically thick, the photosphere necessarily emits as a blackbody, i.e., for distant observers
\begin{equation}
    F_\nu = \Omega B_\nu(T)
    \label{eq:Fnu}
\end{equation}
where $\Omega$ an is the apparent solid angle of the photosphere, $T$ is the temperature at infinity, and $B_\nu$ is the standard Planck function.  The temperature is determined by \autoref{eq:Lpho}, which via the Stefan-Boltzmann law gives
\begin{equation}
    \Lpho=A\sigma T^4, 
    \label{eq:Lpho2}
\end{equation}
where $A$ is the apparent surface area of the photosphere as seen by distant observers and $\sigma$ is the Stefan-Boltzmann constant.  Assuming spherical symmetry,
\begin{equation}
    A = 4\pi R^2
    ~~\text{and}~~
    \Omega = \frac{\pi R^2}{D^2},
\end{equation}
where $R$ is the apparent radius and $D$ is the AGN's distance.

For RIAFs, \autoref{eq:Lbol}, \ref{eq:Lpho}, and \ref{eq:Lpho2} provide a means to express the luminosity and temperature of the putative photosphere in terms of the observed luminosity directly from the accretion flow:
\begin{equation}
    \Lpho = \frac{\eadv}{\eacc} \Lbol
\end{equation}
from which we obtain
\begin{equation}
    T = \left(\frac{\eadv \dot{M} c^2}{\sigma A} \right)^{1/4}
    = \left(\frac{\eadv}{\eacc} \frac{\Lbol}{4\pi \sigma R^2}\right)^{1/4}.
    \label{eq:T}
\end{equation}
We note now a point that will be important later: the temperature is only weakly dependent on the specific efficiencies ($\eacc$ and $\eadv$), AGN luminosity ($\Lbol$), and apparent source size ($R$).  As a consequence, we will find that the $T$ is reasonably narrowly distributed resulting in $F_\nu$ that peak in the extreme UV (EUV) and soft X-rays. 

While $\Lbol$ and $A$ may be estimated in principle, we may simplify \autoref{eq:T} further such that it is a function solely of directly measurable quantities.  The bolometric luminosity is related to the directly measured bolometric flux via $\Lbol=4\pi D^2 \Fbol$.  Similarly, the apparent photosphere radius may be related to its apparent angular radius via $R= D\theta$.  Thus, we find upon substituting these into \autoref{eq:T} that the putative photosphere temperature simplifies to
\begin{equation}
    T = \left( \frac{\eadv}{\eacc} \frac{\Fbol}{\sigma \theta^2}  \right)^{1/4}.
\end{equation}

\subsection{Impact of Cosmological Redshift}
While \autoref{eq:Fnu} and \ref{eq:T} account for the gravitational redshift associated with the compact object, for clarity we have set aside the impact of the cosmological redshift, $z$, for distance AGN.  In practice, we will find that this is unlikely to matter for the targets of interest (all of which have $z<0.1$).  Nevertheless, in the interest of completeness we provide the general expressions here.  In what follows primed quantities reference those in the AGN rest-frame (though still representing a locally distant observer) and unprimed quantities are those observed at Earth.

There are two effects that must be considered.  First, the observation frequency changes, i.e., $\nu = \nu'/(1+z)$.  Closely related, the temperature redshifts in an identical way, i.e., $T = T'/(1+z)$.  These offset, recovering that $B_\nu(T)=B_{\nu'}(T')$.  Second, the object is lensed due to the intervening cosmological expansion, and therefore $\Omega = \pi R'^2/D_A^2$, where $D_A$ is the angular diameter distance.  Therefore,
\begin{equation}
    F_\nu = \pi \frac{R'^2}{D_A^2} B_\nu(T),
\label{eq:Fnu2}
\end{equation}
from which we can immediately obtain the observed bolometric flux, 
\begin{equation}
    \Fbol=\int d\nu F_\nu = \frac{R'^2}{D_A^2} \sigma T^4.
\end{equation} 
The rest-frame bolometric luminosity is related to the $\Fbol$ via the luminosity distance, $D_L=(1+z)^2 D_A$, which accounts for both the cosmological lensing and redshift corrections, giving $\Lbol' = 4\pi D_L^2 \Fbol$.  Therefore, in terms of $\Lbol'$, which can be estimated, and the rest-frame apparent radius of the photosphere, we have
\begin{equation}
    T = 
    \frac{1}{(1+z)} \left( \frac{\eadv}{\eacc} \frac{\Lbol'}{\sigma A'} \right)^{1/4}
    =
    \left( \frac{\eadv}{\eacc} \frac{\Fbol}{\sigma \theta^2} \right)^{1/4},
\end{equation}
where as before we have also expressed $T$ in terms of the locally measured flux and source angular size.  We will drop the primes henceforth, and which frame the quantities are defined in may be identified via context.

\subsection{Summary and Inputs}
\label{sec:2-SI}

For sufficiently low-luminosity sources ($\Lbol$ below 1\% of Eddington), an accretion-powered optically thick photosphere will develop.  This photosphere will thermalize and reach an approximate steady state between its thermal emission and accretion power on timescales that are short in comparison with those relevant for AGNs.  The temperature of this putative photosphere may be estimated given two observational inputs: the observed bolometric flux, $\Fbol$, and apparent angular radius, $\theta$.

In practice, identifying which sources are likely to be in the RIAF state and estimate the putative photosphere temperature (and therefore spectral flux density) from the following quantities,
\begin{enumerate}
    \item Bolometric luminosity of the accretion flow, $\Lbol$.
    \item Black hole candidate mass, $\MBH$.
    \item Apparent radius of the putative photosphere, $R$.
    \item The angular diameter radius, $D_A$.
    \item The cosmological redshift, $z$.
\end{enumerate}
We next turn to identifying how to obtain these from the BASS sample.

\section{BASS Sample}
\label{sec:3}
The BAT AGN Spectroscopic Survey (BASS) is an all-sky survey that traces bright sources above a certain flux threshold in the hard X-ray energy band 14–195 keV, based on the SWIFT/BAT 70-month parent sample survey \citep{Baumgartner2013}. The BASS Data Release 2 (DR2) provides 1449 optical spectra for the 858 Active Galactic Nuclei \citep[AGNs,][]{Koss2022b}. The spectral dataset is high resolution, high sensitivity, and has broad spectral coverage. 
How objects were selected from the BASS sample, their relevant BASS parameters and their associated uncertainties are described below.

\subsection{Source Selection}
We consider only the 104 of the 858 AGNs within the BASS sample with Eddington ratios below 1\%.  These objects are consistent with having luminosity produced within a RIAF \citep{Yuan2014}.  
We further assess the optical depth of their emission regions using a simple one-zone model in which the observed emission is generated via synchrotron emission (see \autoref{app:tau}).  We find that all 104 objects are optically thin at the frequencies relevant for constraining the putative photosphere emission.
Therefore, we expect all of the AGNs in our subsample of the BASS catalog to satisfy the necessary astrophysical conditions for producing an observable accretion powered photosphere.

\startlongtable
\begin{deluxetable*}{l l c c c c l c c c l l}
\tablecaption{Selected source properties for BASS AGN catalog members with Eddington ratios below 1\%.\label{table1}}
\tabletypesize{\scriptsize}
\tablehead{
\colhead{BASS} & \colhead{Common} & \colhead{} & \colhead{RA} & \colhead{Dec} & \colhead{} & \colhead{$\log D_L$} & \colhead{$\log \MBH$} & 
\colhead{$\log \Lbol$\tablenotemark{a}} & \colhead{} & \colhead{$\log \dot{M}$} & \colhead{} \\[-1em]
\colhead{ID} & \colhead{Name} & \colhead{Type} & \colhead{h/m/s} & \colhead{d/m/s} & \colhead{$z$\tablenotemark{b}} & \colhead{(Mpc)} & \colhead{($M_\odot$)} & \colhead{(erg\,s$^{-1}$)} & \colhead{$\lambda_{\rm Edd}$} & \colhead{($M_\odot$\,yr$^{-1}$)} & \colhead{Result\tablenotemark{c}}
}

\startdata
28 & NGC 235A & Sy1.9 & 00 42 52.81 & -23 32 27.77 & 0.02207 & 1.98 $\pm$ 0.1 & 8.49 $\pm$ 0.45 & 44.61 & 0.0087 & -0.1431 & Spec. \\
49 & MCG -7-3-7 & Sy2 & 01 05 26.82 & -42 12 58.38 & 0.03029 & 2.12 $\pm$ 0.1 & 8.39 $\pm$ 0.45 & 44.45 & 0.0076 & -0.3031 & Spec. \\
62 & IC 1657 & Sy2 & 01 14 07.01 & -32 39 03.18 & 0.01169 & 1.68 $\pm$ 0.125 & 7.68 $\pm$ 0.45 & 43.51 & 0.0045 & -1.2431 & Spec. \\
64 & NGC 452 & Sy2 & 01 16 14.80 & +31 02 01.83 & 0.01690 & 1.87 $\pm$ 0.1 & 8.16 $\pm$ 0.45 & 43.68 & 0.0022 & -1.0731 & Src. \\
83 & ESO 353-9 & Sy2 & 01 31 50.40 & -33 07 09.44 & 0.01639 & 1.85 $\pm$ 0.1 & 7.96 $\pm$ 0.45 & 43.82 & 0.0048 & -0.9331 & Src. \\
93 & NGC 678 & Sy2 & 01 49 24.84 & +21 59 50.46 & 0.00949 & 1.54 $\pm$ 0.125 & 8.28 $\pm$ 0.45 & 42.83 & 0.0002 & -1.9231 & Src. \\
96 & MCG-1-5-47 & Sy2 & 01 52 49.04 & -03 26 48.64 & 0.01673 & 1.86 $\pm$ 0.1 & 7.89 $\pm$ 0.45 & 44.02 & 0.0091 & -0.7331 & Exc. \\
103 & LEDA 89913 & Sy2 & 02 02 17.36 & +68 21 45.56 & 0.01184 & 1.59 $\pm$ 0.125 & 7.95 $\pm$ 0.45 & 43.08 & 0.0009 & -1.6731 & Src. \\
104 & LEDA 137972 & Sy1.9 & 02 01 32.35 & +68 24 21.87 & 0.01203 & 1.72 $\pm$ 0.1 & 8.04 $\pm$ 0.45 & 43.24 & 0.0010 & -1.5131 & Src. \\
110 & UGC 1609 & Sy1.9 & 02 07 51.59 & +44 50 38.40 & 0.02167 & 1.97 $\pm$ 0.1 & 7.74 $\pm$ 0.45 & 43.87 & 0.0091 & -0.8831 & Src. \\
112 & NGC 835 & Sy2 & 02 09 20.84 & -10 07 59.08 & 0.01334 & 1.76 $\pm$ 0.1 & 7.75 $\pm$ 0.45 & 43.50 & 0.0038 & -1.2531 & Src. \\
133 & NGC 973 & Sy2 & 02 34 20.11 & +32 30 20.59 & 0.01566 & 1.83 $\pm$ 0.1 & 8.48 $\pm$ 0.45 & 44.08 & 0.0026 & -0.6731 & Exc. \\
140 & NGC 1052 & Sy2 & 02 41 04.80 & -08 15 20.73 & 0.00452 & 1.28 $\pm$ 0.06 & 8.82 $\pm$ 0.45 & 42.93 & 0.0001 & -1.8231 & Exc. \\
146 & ESO 479-31 & Sy1.9 & 02 44 47.72 & -24 30 49.85 & 0.02300 & 2.00 $\pm$ 0.1 & 7.86 $\pm$ 0.45 & 43.84 & 0.0063 & -0.9131 & Src. \\
152 & NGC 1106 & Sy2 & 02 50 40.48 & +41 40 17.76 & 0.01439 & 1.79 $\pm$ 0.1 & 8.33 $\pm$ 0.45 & 44.02 & 0.0033 & -0.7331 & Src. \\
173 & NGC 1275 & Sy1.9 & 03 19 48.16 & +41 30 42.23 & 0.01681 & 1.86 $\pm$ 0.1 & 8.60 $\pm$ 0.5 & 44.55 & 0.0059 & -0.2031 & Src. \\
207 & Fairall 1119 & Sy1 & 04 05 01.70 & -37 11 15.27 & 0.05494 & 2.39 $\pm$ 0.1 & 8.57 $\pm$ 0.35 & 44.72 & 0.0095 & -0.0331 & Exc. \\
216 & NGC 1566 & Sy1 & 04 20 00.39 & -54 56 16.60 & 0.00474 & 1.25 $\pm$ 0.03 & 6.83 $\pm$ 0.5 & 42.73 & 0.0054 & -2.0231 & Not exc. \\
237 & LEDA 86269 & Sy2 & 04 44 09.02 & +28 13 00.76 & 0.01053 & 1.66 $\pm$ 0.125 & 7.98 $\pm$ 0.45 & 43.92 & 0.0058 & -0.8331 & Exc. \\
239 & UGC 3157 & Sy1.9 & 04 46 29.67 & +18 27 39.11 & 0.01567 & 1.83 $\pm$ 0.1 & 8.26 $\pm$ 0.45 & 43.87 & 0.0027 & -0.8831 & Exc. \\
260 & - & Sy1.9 & 05 08 19.71 & +17 21 48.05 & 0.01736 & 1.88 $\pm$ 0.1 & 8.35 $\pm$ 0.45 & 44.06 & 0.0034 & -0.6931 & Exc. \\
263 & Z 307-7 & Sy2 & 05 13 16.42 & +66 27 50.28 & 0.01510 & 1.82 $\pm$ 0.1 & 8.01 $\pm$ 0.45 & 43.60 & 0.0026 & -1.1531 & Src. \\
279 & ESO 553-43 & Sy2 & 05 26 27.27 & -21 17 11.68 & 0.02780 & 2.08 $\pm$ 0.1 & 8.48 $\pm$ 0.45 & 44.25 & 0.0039 & -0.5031 & Exc. \\
302 & - & Sy2 & 05 44 00.09 & -43 25 26.88 & 0.04420 & 2.29 $\pm$ 0.1 & 8.23 $\pm$ 0.45 & 44.39 & 0.0095 & -0.3631 & Src. \\
308 & NGC 2110 & Sy2 & 05 52 11.38 & -07 27 22.47 & 0.00750 & 1.54 $\pm$ 0.125 & 8.78 $\pm$ 0.45 & 44.50 & 0.0035 & -0.2531 & Exc. \\
317 & UGC 3386 & Sy2 & 06 02 37.99 & +65 22 16.44 & 0.01486 & 1.81 $\pm$ 0.1 & 8.38 $\pm$ 0.45 & 43.81 & 0.0018 & -0.9431 & Exc. \\
325 & Mrk 3 & Sy1.9 & 06 15 36.36 & +71 02 14.94 & 0.01382 & 1.78 $\pm$ 0.1 & 8.96 $\pm$ 0.45 & 44.92 & 0.0062 & 0.1669 & Exc. \\
344 & LEDA 549777 & Sy2 & 06 40 38.02 & -43 21 20.56 & 0.06049 & 2.43 $\pm$ 0.1 & 8.88 $\pm$ 0.45 & 45.05 & 0.0098 & 0.2969 & Exc. \\
370 & LEDA 96373 & Sy2 & 07 26 26.34 & -35 54 21.86 & 0.02995 & 2.12 $\pm$ 0.1 & 9.62 $\pm$ 0.45 & 44.50 & 0.0005 & -0.2531 & Exc. \\
383 & Mrk 78 & Sy2 & 07 42 41.70 & +65 10 37.55 & 0.03731 & 2.22 $\pm$ 0.1 & 8.53 $\pm$ 0.45 & 44.43 & 0.0054 & -0.3231 & Src. \\
385 & UGC 3995B & Sy2 & 07 44 09.12 & +29 14 50.86 & 0.01590 & 1.84 $\pm$ 0.1 & 7.97 $\pm$ 0.45 & 43.92 & 0.0060 & -0.8331 & Src. \\
395 & LEDA 2272950 & Sy1.9 & 07 52 44.20 & +45 56 57.48 & 0.05165 & 2.36 $\pm$ 0.1 & 8.66 $\pm$ 0.45 & 44.78 & 0.0089 & 0.0269 & Src. \\
397 & - & Sy2 & 07 56 19.62 & -41 37 42.14 & 0.02114 & 1.96 $\pm$ 0.1 & 7.89 $\pm$ 0.45 & 43.86 & 0.0062 & -0.8931 & Spec. \\
436 & NGC 2655 & Sy2 & 08 55 37.99 & +78 13 23.90 & 0.00485 & 1.39 $\pm$ 0.125 & 8.34 $\pm$ 0.45 & 42.88 & 0.0002 & -1.8731 & Exc. \\
437 & NGC 2712 & Sy2 & 08 59 30.49 & +44 54 50.25 & 0.00675 & 1.49 $\pm$ 0.125 & 6.98 $\pm$ 0.45 & 42.84 & 0.0049 & -1.9131 & Spec. \\
439 & Mrk 18 & Sy1.9 & 09 01 58.40 & +60 09 06.26 & 0.01106 & 1.68 $\pm$ 0.1 & 7.72 $\pm$ 0.45 & 43.36 & 0.0030 & -1.3931 & Src. \\
456 & - & Sy2 & 09 23 53.73 & -31 41 30.99 & 0.04233 & 2.27 $\pm$ 0.1 & 8.80 $\pm$ 0.45 & 44.94 & 0.0091 & 0.1869 & Exc. \\
463 & Z 312-12 & Sy2 & 09 29 37.88 & +62 32 38.63 & 0.02561 & 2.05 $\pm$ 0.1 & 8.16 $\pm$ 0.45 & 43.92 & 0.0039 & -0.8331 & Src. \\
465 & ESO 565-19 & Sy2 & 09 34 43.55 & -21 55 40.03 & 0.01596 & 1.84 $\pm$ 0.1 & 8.45 $\pm$ 0.45 & 44.56 & 0.0085 & -0.1931 & Src. \\
471 & NGC 2992 & Sy1.9 & 09 45 41.94 & -14 19 34.70 & 0.00767 & 1.58 $\pm$ 0.125 & 7.97 $\pm$ 0.45 & 43.48 & 0.0022 & -1.2731 & Exc. \\
498 & ESO 317-38 & Sy2 & 10 29 45.60 & -38 20 54.83 & 0.01506 & 1.81 $\pm$ 0.1 & 7.97 $\pm$ 0.45 & 43.59 & 0.0028 & -1.1631 & Spec. \\
502 & NGC 3281 & Sy2 & 10 31 52.08 & -34 51 13.27 & 0.01113 & 1.68 $\pm$ 0.1 & 8.22 $\pm$ 0.45 & 44.36 & 0.0091 & -0.3931 & Exc. \\
503 & ESO 436-34 & Sy2 & 10 32 44.54 & -28 36 36.54 & 0.01209 & 1.72 $\pm$ 0.1 & 8.12 $\pm$ 0.45 & 43.55 & 0.0018 & -1.2031 & Spec. \\
504 & Z 333-38 & Sy2 & 10 34 23.55 & +73 00 50.44 & 0.02241 & 1.99 $\pm$ 0.1 & 8.30 $\pm$ 0.45 & 44.03 & 0.0035 & -0.7231 & Src. \\
517 & UGC 5881 & Sy2 & 10 46 42.51 & +25 55 53.76 & 0.02046 & 1.95 $\pm$ 0.1 & 8.09 $\pm$ 0.45 & 44.13 & 0.0072 & -0.6231 & Src. \\
521 & LEDA 32573 & Sy2 & 10 51 37.46 & -17 07 29.00 & 0.01856 & 1.91 $\pm$ 0.1 & 8.48 $\pm$ 0.45 & 43.32 & 0.0005 & -1.4331 & Exc. \\
528 & Z 291-28 & Sy2 & 11 05 59.04 & +58 56 45.77 & 0.04775 & 2.33 $\pm$ 0.1 & 8.37 $\pm$ 0.45 & 44.38 & 0.0068 & -0.3731 & Src. \\
543 & Mrk 423 & Sy2 & 11 26 48.52 & +35 15 02.64 & 0.03238 & 2.15 $\pm$ 0.1 & 8.03 $\pm$ 0.45 & 44.13 & 0.0085 & -0.6231 & Src. \\
548 & NGC 3718 & Sy2 & 11 32 34.86 & +53 04 04.70 & 0.00328 & 1.23 $\pm$ 0.125 & 9.05 $\pm$ 0.45 & 42.48 & 0.0000 & -2.2731 & Exc. \\
560 & NGC 3786 & Sy1.9 & 11 39 42.51 & +31 54 33.94 & 0.00892 & 1.64 $\pm$ 0.125 & 7.82 $\pm$ 0.45 & 43.42 & 0.0026 & -1.3331 & Src. \\
568 & UGC 6732 & Sy2 & 11 45 33.13 & +58 58 41.36 & 0.00729 & 1.50 $\pm$ 0.1 & 6.96 $\pm$ 0.45 & 43.10 & 0.0093 & -1.6531 & Spec. \\
573 & MCG +5-28-32 & Sy2 & 11 48 45.93 & +29 38 28.36 & 0.02283 & 2.00 $\pm$ 0.1 & 8.62 $\pm$ 0.45 & 44.30 & 0.0032 & -0.4531 & Src. \\
579 & NGC 3998 & Sy1.9 & 11 57 56.13 & +55 27 13.01 & 0.00357 & 1.18 $\pm$ 0.06 & 8.93 $\pm$ 0.5 & 42.48 & 0.0000 & -2.2731 & Exc. \\
580 & IC 751 & Sy2 & 11 58 52.62 & +42 34 13.17 & 0.03119 & 2.14 $\pm$ 0.1 & 8.57 $\pm$ 0.45 & 44.38 & 0.0044 & -0.3731 & Exc. \\
590 & NGC 4102 & Sy2 & 12 06 23.05 & +52 42 39.71 & 0.00236 & 1.29 $\pm$ 0.125 & 7.84 $\pm$ 0.45 & 43.36 & 0.0022 & -1.3931 & Src. \\
593 & NGC 4138 & Sy2 & 12 09 29.80 & +43 41 07.11 & 0.00319 & 1.14 $\pm$ 0.06 & 7.71 $\pm$ 0.45 & 42.69 & 0.0006 & -2.0631 & Src. \\
599 & NGC 4180 & Sy2 & 12 13 03.07 & +07 02 20.10 & 0.00653 & 1.63 $\pm$ 0.125 & 7.63 $\pm$ 0.45 & 43.59 & 0.0062 & -1.1631 & Src. \\
607 & NGC 4235 & Sy1 & 12 17 09.88 & +07 11 29.79 & 0.00793 & 1.42 $\pm$ 0.06 & 7.28 $\pm$ 0.35 & 43.26 & 0.0063 & -1.4931 & Src. \\
609 & NGC 4258 & Sy1.9 & 12 18 57.50 & +47 18 14.45 & 0.00169 & 0.89 $\pm$ 0.03 & 7.56 $\pm$ 0.5 & 41.99 & 0.0002 & -2.7631 & Src. \\
621 & NGC 4500 & Sy2 & 12 31 22.19 & +57 57 52.63 & 0.01036 & 1.54 $\pm$ 0.125 & 7.73 $\pm$ 0.45 & 43.10 & 0.0016 & -1.6531 & Src. \\
633 & NGC 4619 & Sy1.9 & 12 41 44.55 & +35 03 45.78 & 0.02295 & 2.00 $\pm$ 0.1 & 7.83 $\pm$ 0.45 & 43.75 & 0.0056 & -1.0031 & Spec. \\
638 & NGC 4686 & Sy2 & 12 46 39.89 & +54 32 03.43 & 0.01660 & 1.86 $\pm$ 0.1 & 9.07 $\pm$ 0.45 & 44.19 & 0.0009 & -0.5631 & Exc. \\
648 & - & Sy2 & 13 00 05.35 & +16 32 14.83 & 0.07998 & 2.56 $\pm$ 0.1 & 9.05 $\pm$ 0.45 & 45.15 & 0.0083 & 0.3969 & Src. \\
653 & NGC 4941 & Sy2 & 13 04 13.10 & -05 33 05.74 & 0.00388 & 1.31 $\pm$ 0.125 & 7.00 $\pm$ 0.45 & 42.89 & 0.0051 & -1.8631 & Src. \\
654 & NGC 4939 & Sy2 & 13 04 14.32 & -10 20 22.21 & 0.01054 & 1.62 $\pm$ 0.125 & 7.75 $\pm$ 0.45 & 43.61 & 0.0049 & -1.1431 & Exc. \\
662 & - & Sy2 & 13 10 57.26 & +08 37 38.32 & 0.05273 & 2.37 $\pm$ 0.1 & 8.94 $\pm$ 0.45 & 44.70 & 0.0038 & -0.0531 & Src. \\
665 & NGC 5033 & Sy1.9 & 13 13 27.48 & +36 35 37.96 & 0.00276 & 1.28 $\pm$ 0.125 & 7.75 $\pm$ 0.45 & 42.22 & 0.0002 & -2.5331 & Src. \\
666 & Mrk 248 & Sy2 & 13 15 17.32 & +44 24 25.88 & 0.03552 & 2.19 $\pm$ 0.1 & 8.62 $\pm$ 0.45 & 44.55 & 0.0058 & -0.2031 & Exc. \\
669 & LEDA 46599 & Sy1.9 & 13 20 59.58 & +08 58 42.12 & 0.03256 & 2.16 $\pm$ 0.1 & 8.40 $\pm$ 0.45 & 44.44 & 0.0072 & -0.3131 & Src. \\
670 & MCG -3-34-64 & Sy1.9 & 13 22 24.46 & -16 43 42.40 & 0.01673 & 1.86 $\pm$ 0.1 & 8.37 $\pm$ 0.45 & 44.27 & 0.0052 & -0.4831 & Src. \\
671 & Cen A & Sy2 & 13 25 27.62 & -43 01 09.19 & 0.00188 & 0.57 $\pm$ 0.03 & 7.77 $\pm$ 0.5 & 43.23 & 0.0019 & -1.5231 & Exc. \\
674 & ESO 509-38 & Sy1 & 13 31 13.83 & -25 24 09.92 & 0.02607 & 2.06 $\pm$ 0.1 & 8.32 $\pm$ 0.35 & 44.14 & 0.0044 & -0.6131 & Exc. \\
676 & ESO 21-4 & Sy2 & 13 32 40.62 & -77 50 40.52 & 0.00963 & 1.62 $\pm$ 0.1 & 8.25 $\pm$ 0.45 & 43.53 & 0.0013 & -1.2231 & Src. \\
679 & NGC 5231 & Sy2 & 13 35 48.25 & +02 59 56.16 & 0.02156 & 1.97 $\pm$ 0.1 & 7.99 $\pm$ 0.45 & 44.00 & 0.0068 & -0.7531 & Src. \\
682 & NGC 5252 & Sy2 & 13 38 15.88 & +04 32 33.58 & 0.02297 & 2.00 $\pm$ 0.1 & 9.00 $\pm$ 0.5 & 44.95 & 0.0059 & 0.1969 & Exc. \\
684 & NGC 5283 & Sy2 & 13 41 05.75 & +67 40 20.01 & 0.01036 & 1.71 $\pm$ 0.125 & 7.55 $\pm$ 0.45 & 43.29 & 0.0037 & -1.4631 & Spec. \\
685 & Mrk 268 & Sy1.9 & 13 41 11.14 & +30 22 41.27 & 0.04085 & 2.26 $\pm$ 0.1 & 8.63 $\pm$ 0.45 & 44.79 & 0.0098 & 0.0369 & Src. \\
686 & NGC 5273 & Sy1 & 13 42 08.38 & +35 39 15.48 & 0.00361 & 1.22 $\pm$ 0.06 & 6.66 $\pm$ 0.5 & 42.41 & 0.0038 & -2.3431 & Spec. \\
688 & NGC 5290 & Sy2 & 13 45 19.17 & +41 42 44.54 & 0.00857 & 1.54 $\pm$ 0.125 & 7.83 $\pm$ 0.45 & 43.25 & 0.0017 & -1.5031 & Src. \\
692 & 4U 1344-60 & Sy1.9 & 13 47 36.01 & -60 37 03.87 & 0.01283 & 1.74 $\pm$ 0.1 & 9.09 $\pm$ 0.35 & 44.42 & 0.0014 & -0.3331 & Exc. \\
739 & NGC 5728 & Sy1.9 & 14 42 23.87 & -17 15 10.86 & 0.01032 & 1.57 $\pm$ 0.125 & 8.25 $\pm$ 0.45 & 44.14 & 0.0052 & -0.6131 & Exc. \\
746 & LEDA 3079667 & Sy1.9 & 14 51 33.14 & -55 40 38.32 & 0.01809 & 1.90 $\pm$ 0.1 & 8.33 $\pm$ 0.45 & 44.32 & 0.0066 & -0.4331 & Exc. \\
766 & NGC 5899 & Sy2 & 15 15 03.25 & +42 02 59.35 & 0.00860 & 1.65 $\pm$ 0.125 & 7.96 $\pm$ 0.45 & 43.56 & 0.0026 & -1.1931 & Exc. \\
767 & Z 319-7 & Sy2 & 15 19 33.67 & +65 35 58.58 & 0.04408 & 2.29 $\pm$ 0.1 & 8.70 $\pm$ 0.45 & 44.63 & 0.0058 & -0.1231 & Exc. \\
778 & LEDA 2730634 & Sy2 & 15 46 24.33 & +69 29 10.24 & 0.03764 & 2.22 $\pm$ 0.1 & 8.51 $\pm$ 0.45 & 44.56 & 0.0074 & -0.1931 & Src. \\
804 & Z 367-9 & Sy2 & 16 19 19.31 & +81 02 47.39 & 0.02392 & 2.02 $\pm$ 0.1 & 8.34 $\pm$ 0.45 & 44.32 & 0.0063 & -0.4331 & Exc. \\
823 & ESO 137-34 & Sy2 & 16 35 14.00 & -58 04 47.87 & 0.00876 & 1.53 $\pm$ 0.125 & 7.48 $\pm$ 0.45 & 43.50 & 0.0069 & -1.2531 & Src. \\
838 & NGC 6221 & Sy2 & 16 52 46.33 & -59 13 01.03 & 0.00411 & 1.08 $\pm$ 0.125 & 6.72 $\pm$ 0.45 & 42.39 & 0.0032 & -2.3631 & Spec. \\
839 & UGC 10593 & Sy2 & 16 52 18.89 & +55 54 20.08 & 0.02936 & 2.11 $\pm$ 0.1 & 8.43 $\pm$ 0.45 & 44.33 & 0.0052 & -0.4231 & Src. \\
841 & NGC 6240 & Sy2 & 16 52 58.90 & +02 24 03.66 & 0.02472 & 2.03 $\pm$ 0.1 & 9.20 $\pm$ 0.5 & 45.04 & 0.0046 & 0.2869 & Exc. \\
942 & NGC 6552 & Sy2 & 18 00 07.27 & +66 36 54.33 & 0.02682 & 2.07 $\pm$ 0.1 & 8.51 $\pm$ 0.45 & 44.51 & 0.0066 & -0.2431 & Src. \\
981 & CGMW5-04382 & Sy1 & 18 30 50.63 & +09 28 42.02 & 0.01927 & 1.92 $\pm$ 0.1 & 7.43 $\pm$ 0.35 & 43.36 & 0.0056 & -1.3931 & Src. \\
1020 & UGC 11397 & Sy2 & 19 03 49.14 & +33 50 41.18 & 0.01512 & 1.82 $\pm$ 0.1 & 7.71 $\pm$ 0.45 & 43.84 & 0.0091 & -0.9131 & Src. \\
1077 & NGC 6921 & Sy2 & 20 28 28.89 & +25 43 24.18 & 0.01406 & 1.78 $\pm$ 0.1 & 8.72 $\pm$ 0.45 & 43.94 & 0.0011 & -0.8131 & Exc. \\
1092 & IC 5063 & Sy2 & 20 52 02.36 & -57 04 07.52 & 0.01127 & 1.66 $\pm$ 0.125 & 8.24 $\pm$ 0.45 & 44.19 & 0.0060 & -0.5631 & Exc. \\
1135 & NGC 7172 & Sy2 & 22 02 01.89 & -31 52 10.53 & 0.00851 & 1.53 $\pm$ 0.125 & 8.15 $\pm$ 0.45 & 44.08 & 0.0056 & -0.6731 & Exc. \\
1142 & NGC 7213 & Sy1 & 22 09 16.21 & -47 10 00.11 & 0.00477 & 1.34 $\pm$ 0.125 & 7.13 $\pm$ 0.35 & 43.22 & 0.0083 & -1.5331 & Src. \\
1152 & UGC 12040 & Sy1.9 & 22 27 05.77 & +36 21 41.70 & 0.02083 & 1.96 $\pm$ 0.1 & 8.05 $\pm$ 0.45 & 44.10 & 0.0074 & -0.6531 & Src. \\
1173 & UGC 12243 & Sy2 & 22 54 43.52 & +11 42 50.79 & 0.02851 & 2.10 $\pm$ 0.1 & 8.12 $\pm$ 0.45 & 44.17 & 0.0076 & -0.5831 & Src. \\
1177 & UGC 12282 & Sy1.9 & 22 58 55.31 & +40 55 56.39 & 0.01776 & 1.89 $\pm$ 0.1 & 8.60 $\pm$ 0.45 & 44.12 & 0.0022 & -0.6331 & Exc. \\
1180 & NGC 7465 & Sy2 & 23 02 00.96 & +15 57 53.18 & 0.00634 & 1.43 $\pm$ 0.125 & 7.02 $\pm$ 0.45 & 42.92 & 0.0052 & -1.8331 & Spec. \\
1184 & NGC 7479 & Sy2 & 23 04 56.63 & +12 19 22.66 & 0.00710 & 1.57 $\pm$ 0.125 & 7.58 $\pm$ 0.45 & 43.41 & 0.0045 & -1.3431 & Src. \\
841 & NGC 6240N* & Sy2 & 16 52 58.90 & +02 24 03.66 & 0.02472 & 2.03 $\pm$ 0.1 & 8.83 $\pm$ 0.45 & 44.62 & 0.0041 & -0.1331 & Exc. \\
112 & NGC 833* & Sy2 & 02 09 20.80 & -10 07 59.16 & 0.013548 & 1.77 $\pm$ 0.1 & 8.31 $\pm$ 0.45 & 43.19 & 0.0005 & -1.5631 & Src. \\
\enddata
\tablenotetext{*}{NGC 6240N (ID:841) and NGC 833 (ID:112) are dual AGNs.}
\tablenotetext{b}{All redshifts have an uncertainty of 0.0005.}
\tablenotetext{a}{All bolometric luminosities have an uncertainty of 0.2 dex.}
\tablenotetext{c}{Result classifications are abbreviated as follows: Exc. = Excluded, Src. = Marginal source, Spec. = Marginal spectrum, and Not exc. = Not excluded.}
\end{deluxetable*}

\subsection{Distance, Redshift}
\label{sec:3-Dz}
The AGN host galaxy distances are divided into two categories: redshift-independent and redshift-dependent. The redshift-independent distances are luminosity distances for nearby AGNs at $<50$~Mpc that rely on different methods to get an estimate since their peculiar velocities can be larger than the Hubble flow of the host galaxy. These distance estimates are obtained using high-quality measurements such as Tip of the Red Giant Branch (TRGB) or Surface Brightness Fluctuations (SBF) from EDD database, or lower-quality measurements such as Tully-Fisher from NED or CF-3 databases (see \cite{Leroy2019}, \cite{Koss2022} for more details). The uncertainty in these methods is as follows: for TRGB, it is 0.03 dex, SBF is 0.06 dex, lower quality measurements are 0.125 dex.\footnote{These uncertainties are log-normal in the distance and describe the 1-$\sigma$ errors in $\log D$.  One dex uncertainty corresponds to an order of magnitude.}
In our sample, 36 AGNs are considered to be redshift-independent. 
The rest of galaxies in the sample have their distances estimated from their redshifts, meaning that they assume that they trace the Hubble flow and ignore potential peculiar velocities.
Therefore, an uncertainty of 0.1 dex is adopted for these redshift-based 
distances \citep{Leroy2019}. 

Most of the redshifts reported in BASS are spectroscopic, predominantly
specifically focusing on [O~III]~$\lambda5007$ emission line. This line occurs in low density regions like the narrow emission line region (NLR). When line fitting was unavailable, BASS DR1 data were used to retrieve the missing redshifts. Most of the redshifts and their associated uncertainties are reported in SIMBAD. For the sources where the SIMBAD redshifts were consistent with the ones in the BASS catalog, the SIMBAD uncertainties were adopted, which corresponds to $\sigma_z = 0.0005$. For the sources (4/104 AGNs) that did not have reported uncertainties in SIMBAD, the same uncertainty was adopted for consistency. 

\begin{figure}[ht]
\centering
\includegraphics[width=\columnwidth]{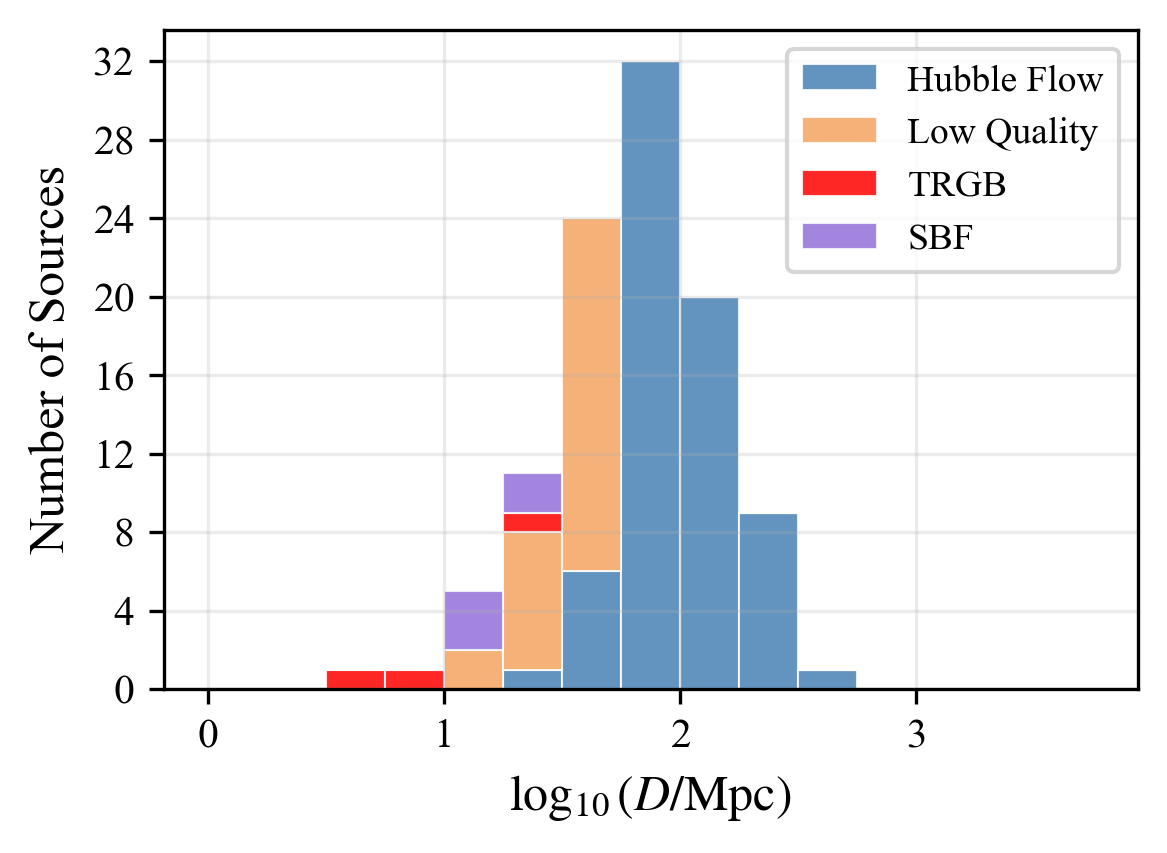}
\includegraphics[width=\columnwidth]{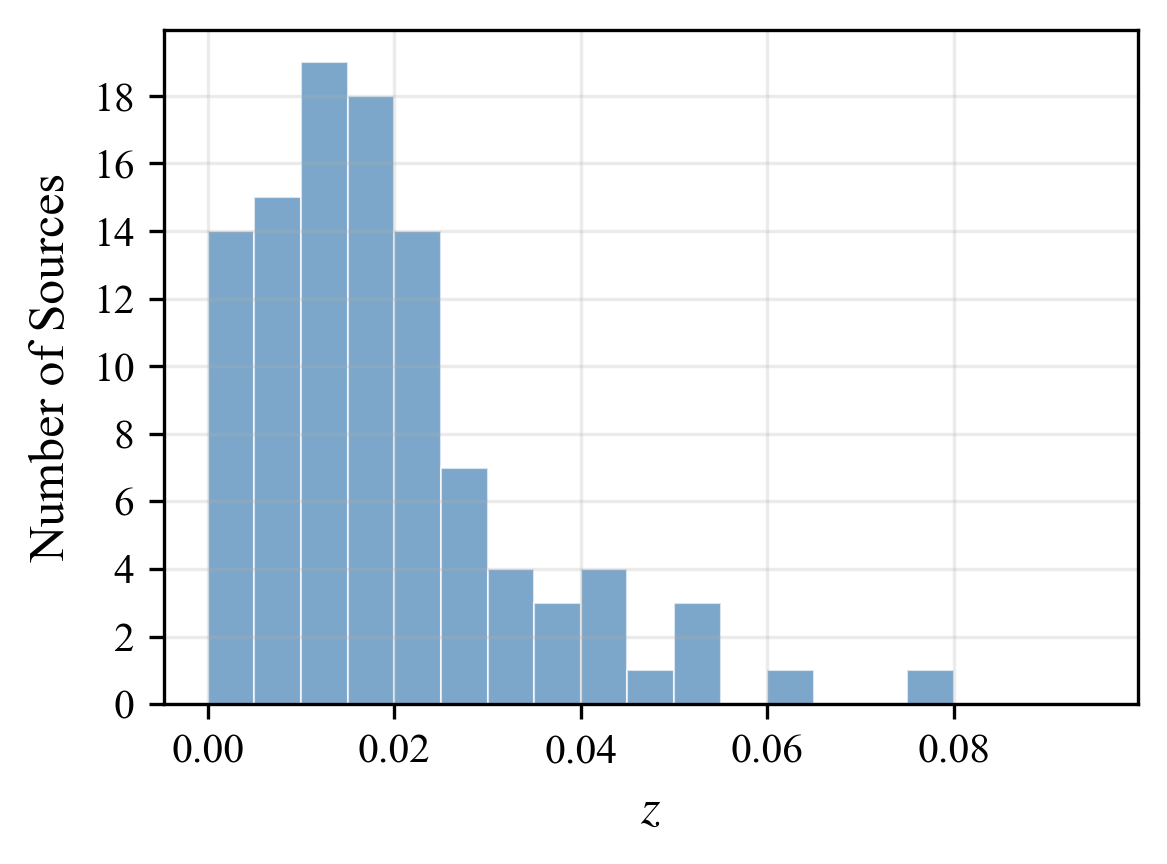}
\caption{Top: The histogram of logarithmic distances reported in Mpc, stacked and color coded by the corresponding measurement methodology. Bottom: the histogram of redshifts reported.}
\label{Fig:z and D hist}
\end{figure}

The distributions of distances and redshifts of our subsample of the BASS catalog are shown in top and bottom panels of \autoref{Fig:z and D hist}, respectively.  Most of our sources are at low redshift; all are below $z=0.1$ and the majority have $z<0.025$.  Correspondingly, most of our sources are closer than $300~\Mpc$.  Consequently, we do not believe that the cosmological redshift is likely to have a significant impact on our analysis.

\subsection{Black Hole Candidate Mass}
\label{sec:3-M} 
The black hole mass estimates reported in the BASS catalog, here associated with the putative horizonless object, were obtained by a variety of methods.  
Where possible, pre-existing estimates from the literature or virial mass estimates using the widths of Balmer lines ($\mathrm{H}\alpha$ and $\mathrm{H}\beta$) arising in the broad-line region (BLR) are used.  Failing those, the mass is estimate from the stellar velocity dispersion and the 
$M_{\text{BH}}\text{--}\sigma_*$ relation.

\begin{figure}[ht]
    \centering
    \includegraphics[width=\columnwidth]{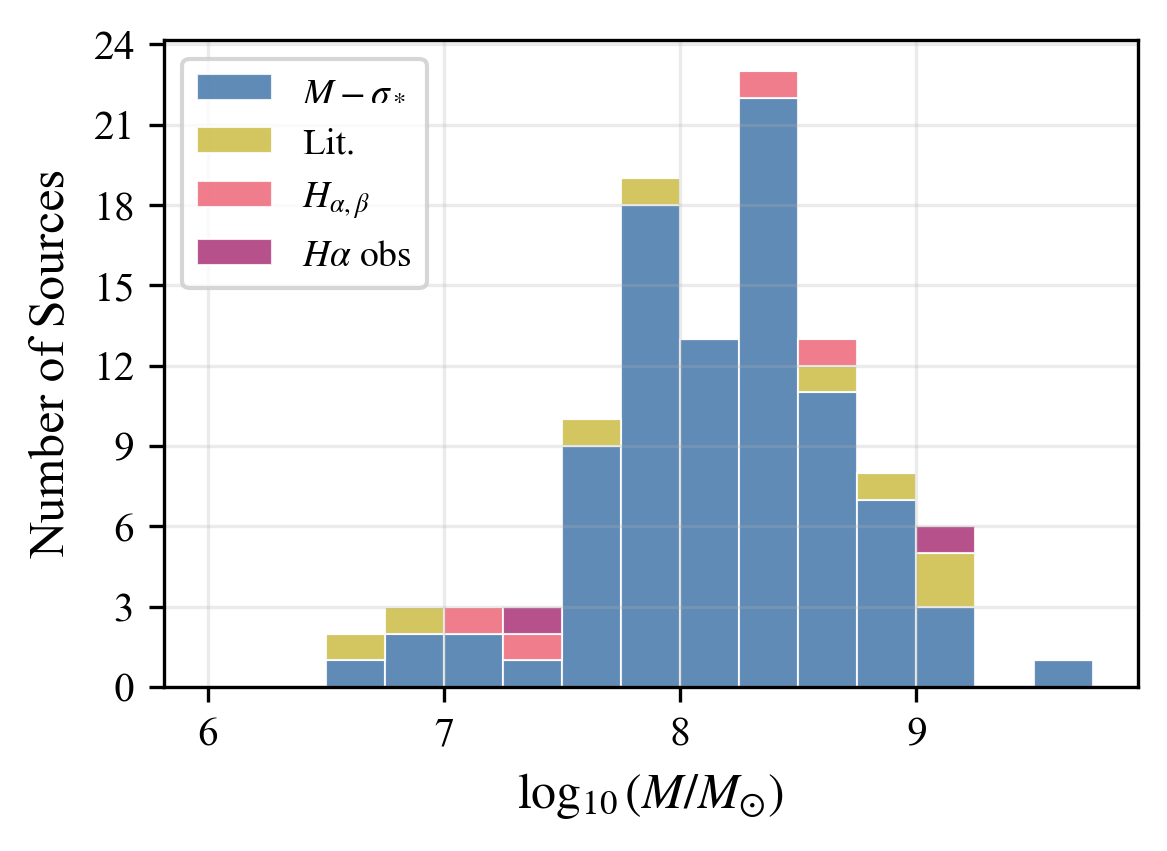}
    \caption{Distribution of logarithmic black hole masses, $\log_{10}(M / M_{\odot})$, stacked and color-coded by the corresponding estimation method.}
    \label{Fig:M hist}
\end{figure}

The distribution of black hole mass estimates for our subsample, and their associated origins, is shown in \autoref{Fig:M hist}.
Most masses lie between $3\times10^7\,M_\odot$ and $10^9\,M_\odot$, indicating that our sample is biased towards higher masses.  The vast majority of our sample has masses obtained from the $M_{\text{BH}}\text{--}\sigma_*$ relation, indicating that the bulk of our sample are obscured type 2 AGN.  

The uncertainty in black hole mass estimates is typically between 0.3 and 0.5 dex, depending on the method being used. If an uncertainty range was given for a specific method, the mean of that range was taken. The $M_{\text{BH}}\text{--}\sigma_*$ relation based estimates have an uncertainty of 0.45 dex \citep{Koss2022}, the single epoch broad line estimates have an uncertainty of 0.35 dex \citep{Koss2022b}, and an uncertainty of 0.5 dex was adopted for mass estimates obtained from the literature. 

\subsection{X-ray Luminosity, Bolometric Luminosity}
\label{sec:3-L}

\begin{figure}[ht]
\centering
\includegraphics[width=\columnwidth]{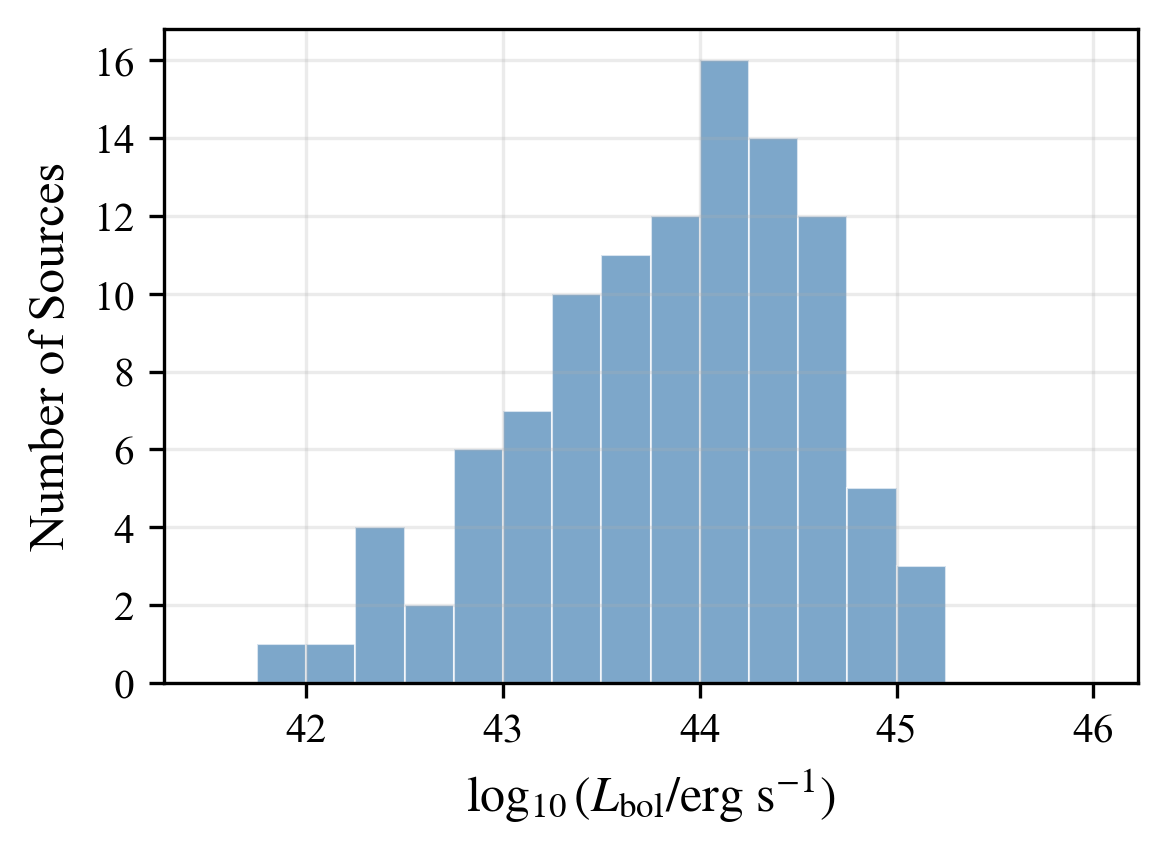}
\caption{The histogram of bolometric luminosities, $\Lbol$, in units of erg s$^{-1}$.}
\label{Fig:L hist}
\end{figure}

The bolometric luminosities ($\Lbol$) reported in the BASS catalog were derived from the intrinsic X-ray luminosities in the 14-150 keV range, with a bolometric correction factor of 8, with an estimated uncertainty of 0.2 dex \citep{Koss2022b}.
We note that by virtue of being estimated from the hard X-rays, $\Lbol$ does not include the putative surface emission discussed in \autoref{sec:2}, which we will see peaks at significantly lower energies.
As shown in \autoref{Fig:L hist}, the $\Lbol$ in our sample ranges from approximately $10^{43}\,\erg\,\s^{-1}$ to $3\times10^{44}\,\erg\,\s^{-1}$, meaning that our AGNs are roughly 1.5 orders of magnitude less luminous on average than those in the complete BASS catalog.  This is a natural consequence of selecting underluminous objects.

\subsection{Spectra}
\label{sec:3-spec}
The flux-calibrated optical spectra were obtained from the BASS DR2. Each spectral file is limited only to wavelength, flux, and the intrinsic error in flux if applicable. The spectral range spans from 3200–10000 \AA.  
 
Instrument-specific masking was applied when it was necessary. For spectra obtained with the Palomar telescope, the wavelength region between 5400 and 5650~\AA\ was masked due to loss of sensitivity. For Magellan, wavelengths above 8280~\AA\ were masked to avoid contamination from significant fringing effects \citep{Koss2022}. The rest of the instruments were not reported to have any sensitivity issues.

Because we are interested in the continuum, we average the spectra in equal logarithmic bins in frequency with widths of 0.03 dex. The stochastic uncertainty of the flux in each frequency bin are estimated by the standard deviation of the spectral flux densities within it. Only points with ${\rm S/N}>1$ are kept.  Finally, a systematic uncertainty in the absolute flux calibration of 30\% is added in quadrature to the stochastic errors \citep{BASS_DR2_README}.

\begin{figure}
    \centering
    \includegraphics[width=\columnwidth]{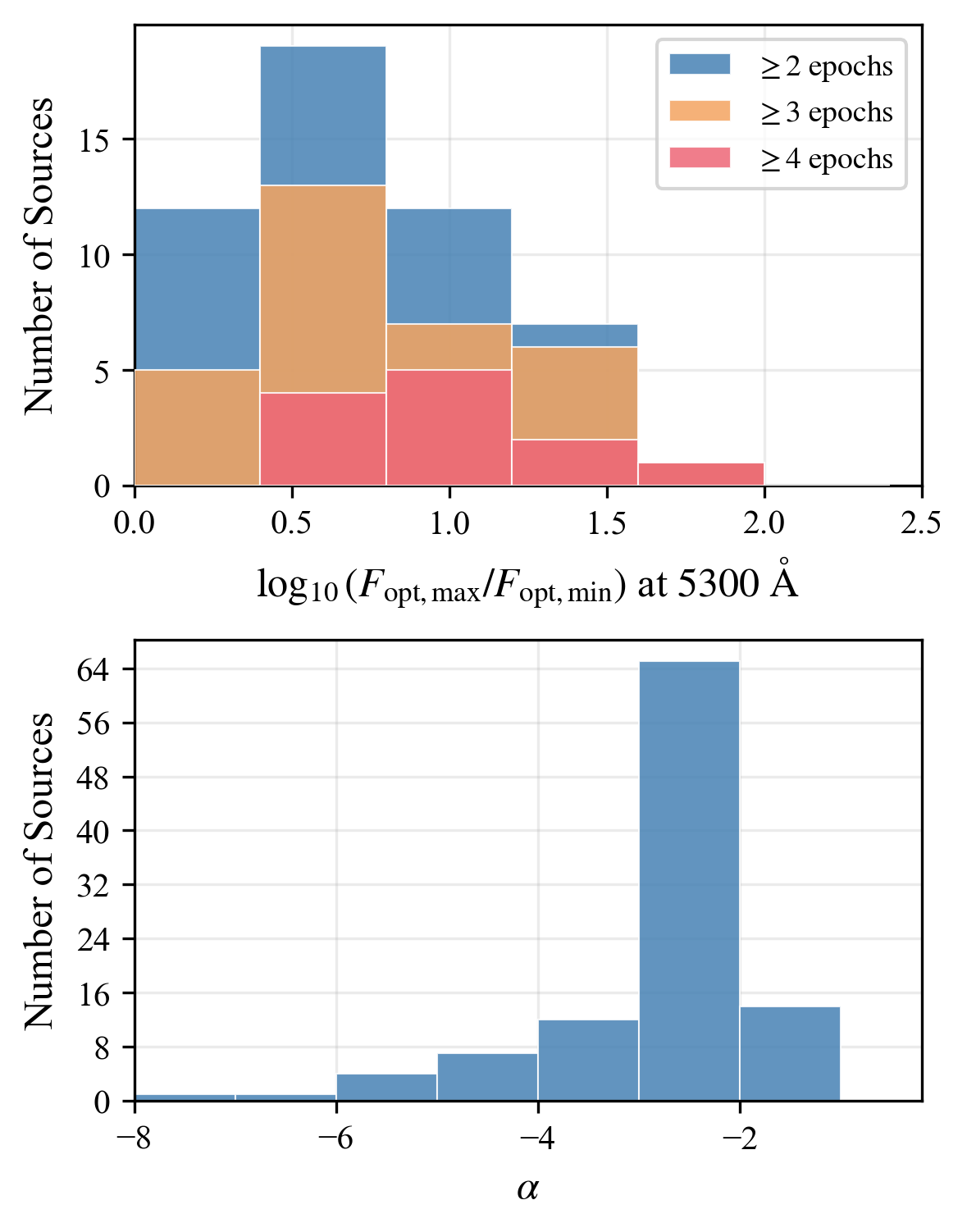}
    \caption{Top: Fractional optical variability ($5300~\mAA$) for the 53 objects in our sample with more than one observation epoch.  Subsets with more than 3 or 4 observations epochs are also shown. Bottom: Spectral index distribution for all of the sources in our sample.}
    \label{fig:fractional_variability}
\end{figure}

The optical spectra in our sample are strongly variable. The top panel of \autoref{fig:fractional_variability} shows the distribution of the maximum relative variation at $5300~\mAA$ for objects with more than one spectral epoch available in the BASS catalog. The median fractional variability is 0.7~dex, with the upper quartile exceeding 1.0~dex; i.e., a quarter of our subsample has optical spectra that very by more than an order of magnitude. The fractional variability increases further only modestly with the number of observation epochs.

The optical spectra also are strongly decreasing with frequency. The distribution of spectral indexes, defined by $F_\nu\propto\nu^{\alpha}$, is shown in the bottom panel of \autoref{fig:fractional_variability}. For no source in our subsample of the BASS catalog do we find $\alpha>0$. The maximum spectral index is $-1.6$, with the median $\alpha=-2.4$. Some evidence for spectral softening with increasing frequency is present, but we do not pursue that further.   

\subsection{Accretion Powered Photosphere Estimates}
\label{sec:3-APPE}

The source properties provided by the BASS catalog as described in \autoref{sec:3-Dz}-\autoref{sec:3-L} provide four of the five necessary inputs in \autoref{sec:2-SI} to estimate the emission from a putative accretion powered photosphere.  Specifically, BASS provides directly estimates of $\Lbol$, $M$, $z$. 

BASS provides luminosity distances ($D_L$), which may be converted into $D_A$ via
\begin{equation}\label{eq:D_A}
    D_A = \frac{D_L}{(1+z)^2}.
\end{equation}
When $z\ll1$, this is only a small correction and therefore well within the uncertainty in the distance estimates themselves for sources in our sample (see \autoref{Fig:z and D hist}.  Therefore, we adopt the BASS distance estimates directly.

The only necessary parameter left unconstrained is the photosphere size, $R$. 
For a handful of objects, it is possible to obtain strong constraints of the size of the emission region by direct imaging \citep{M87_PaperVI,SgpaperVI}.  Indirect upper limits from lensing are also possible \citep{Morgan2010,Blackburne2011}
More commonly, indirect upper limits may be generated from the timescale of large-amplitude brightness fluctuations 
\citep[][]{Gaidos1996,Celotti1998} .

We were unable to find published size constraints for sources in our subsample of the BASS catalog. In \autoref{App:Size_estimate} we explore the viability of using archival X-ray data (HEAO-1, EXOSAT, Einstein, Chandra) to set size constraints on the inner accretion flow size of Cen A via the flux power spectrum, without success.
Therefore, we assume that the photosphere has a characteristic apparent size comparable to that of the shadow of a black hole, i.e., $R\approx\sqrt{27}GM/c^2$.  Were this size is much larger, i.e., an accretion-powered photosphere exists and is much larger than the photon sphere of the corresponding mass black hole, we would expect it to significantly impact the accretion process and substantially modify the hard X-ray emission itself. Nevertheless, this is a key assumption, and one to which we will return in the next section.

\section{Analysis and Results}
\label{sec:4}

With the source parameters provided by the BASS sample listed in \autoref{sec:3-APPE}, and therefore the anticipated spectrum of a putative accretion-powered photosphere, and the BASS source spectra in \autoref{sec:3-spec}, it is possible to directly constrain the existence of photosphere emission.  This provides a clear test of a generic feature of black hole foils.  That is, we compare the observed optical spectra of the 104 AGNs from our subsample of the BASS AGN catalog to their accretion-powered photosphere spectrum models and assess if such a feature may be present instead of an event horizon. 

\begin{figure}[t]
    \centering
    \includegraphics[width=\columnwidth]{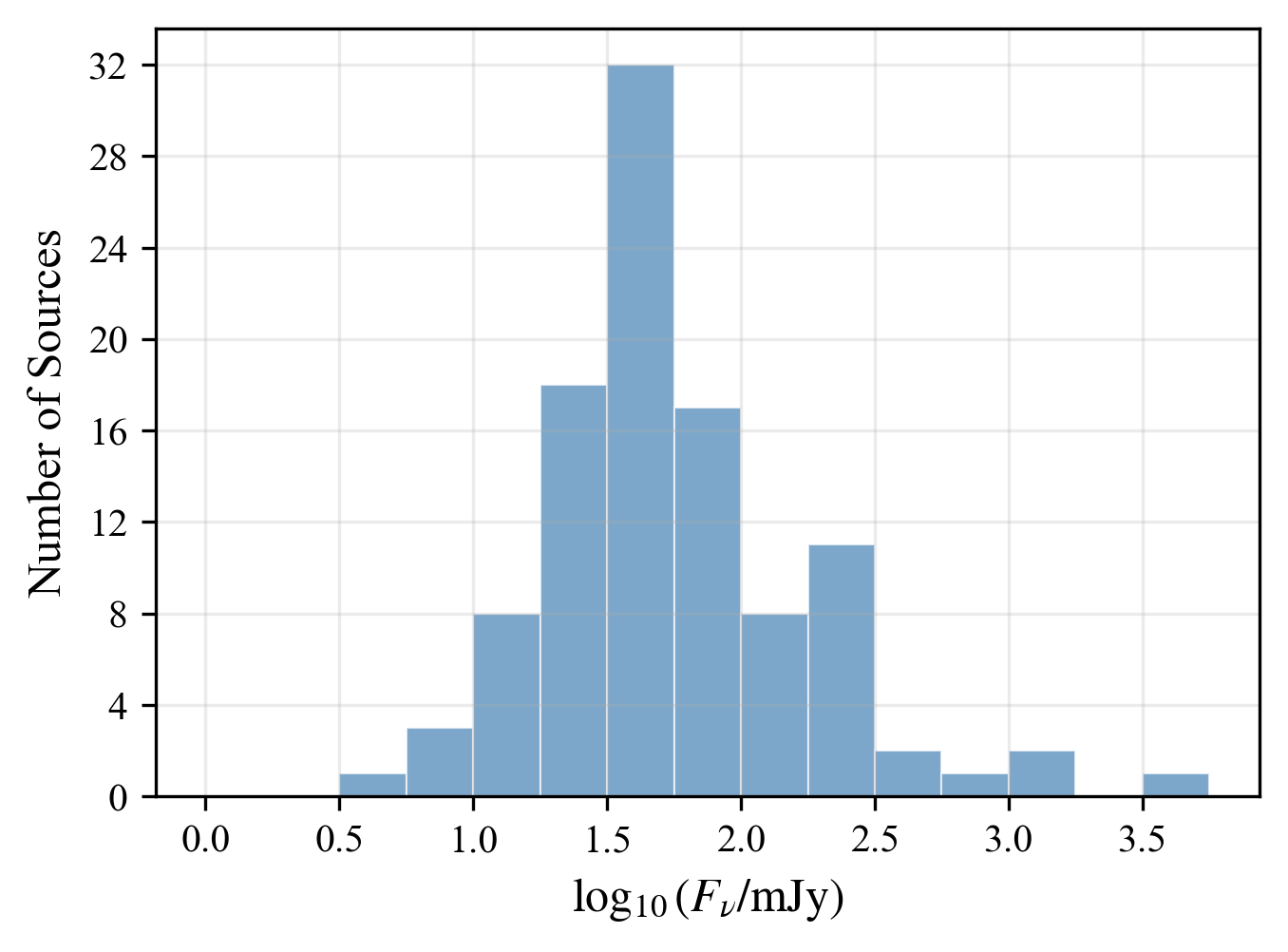}\\
    \includegraphics[width=\columnwidth]{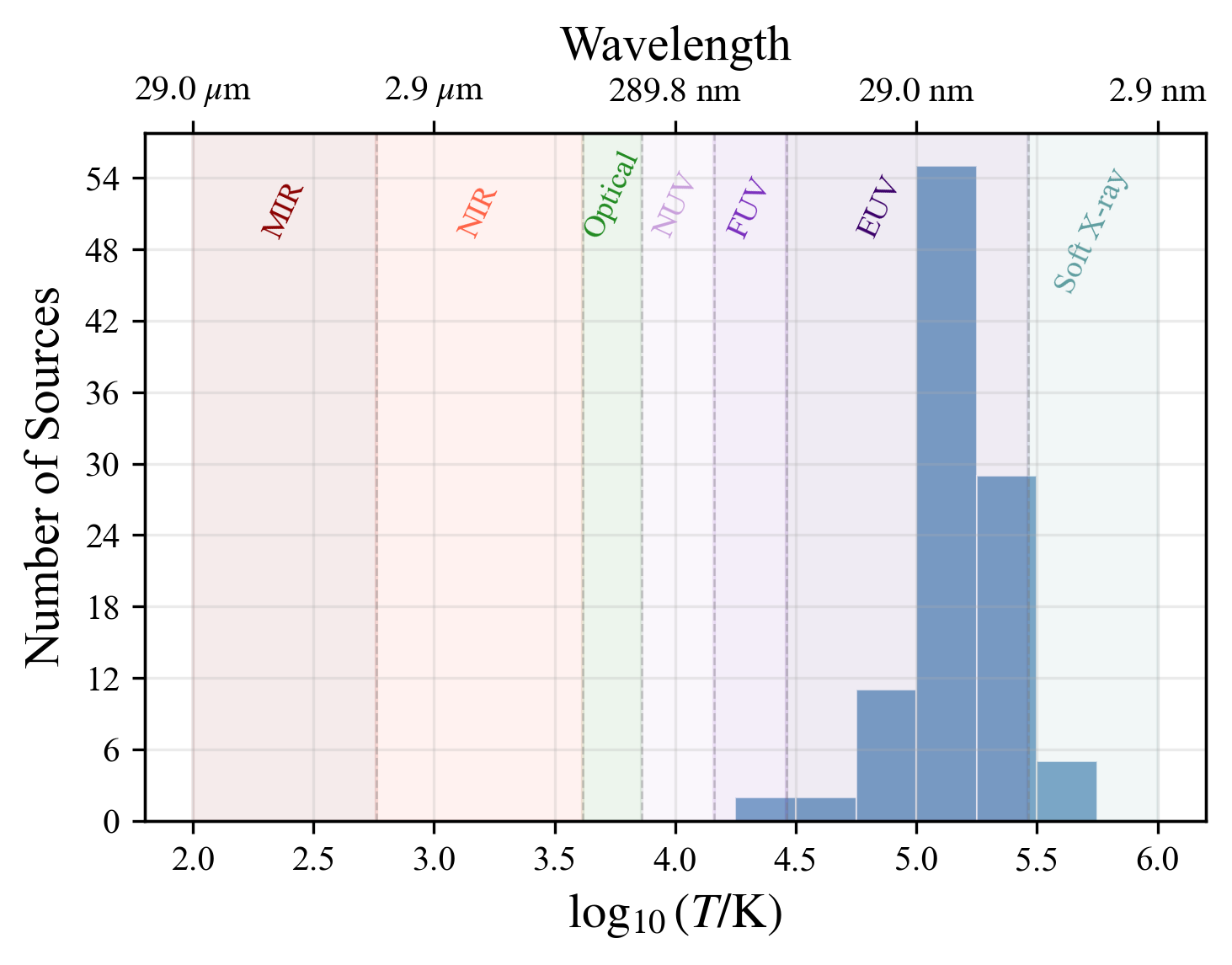}
    \caption{Top: The histogram of the peak predicted fluxes of each AGN in mJy. Bottom: The histogram of predicted surface temperatures of each AGN. The upper axis shows the corresponding wavelengths at each order of magnitude in temperature. The shaded regions span from Mid-Infrared to Soft X-ray range, while the predicted surface temperatures range from the Far-Ultraviolet to Soft X-ray range}
    \label{Fig:Peak Flux and T}
\end{figure}

Across all sources in our sample, the predicted flux density peaks within the UV and X-ray regimes, as seen in \autoref{Fig:Peak Flux and T}. The narrowness of the distribution arises from the weak dependence of the photosphere temperature on the intrinsic source properties.  That is, because $T\propto \Fbol^{1/4} \theta^{-1/2}$, despite being distributed over 1.5 dex in $\Fbol$ and roughly a decade in $\theta$, the temperature of the almost all of our sources lies between $1-3\times10^5~\K$.  As a direct consequence, the peak of the associated Planck function typically occurs in the extreme ultraviolet (EUV).

For the purpose of comparing with the spectra in our subsample of the BASS catalog, the narrow range of surface temperatures has two important consequences. First, the frequency of the anticipated photosphere spectra is $\nu_{\rm peak}\approx 5.9\times10^{15} (T/10^{15}~\K)~{\rm Hz}$, and thus are well within the Rayleigh-Jeans regime at optical frequencies.  That is, where dominated by the photosphere emission we expect an inverted spectral slope with a spectral index of $\approx 2$.  This stands in stark contrast to the typical $\alpha=-2.4$, as discussed in \autoref{sec:3-spec}.

Second, and more important, is the fact that the spectral peak of the photosphere emission lies well above the optical range all but completely mitigates the impact of an uncertain source size. While increasing the source size decreases the photosphere temperature (see \autoref{eq:T}), at a fixed frequency $\nu\ll\nu_{\rm peak}$ the flux scales as $F_\nu \propto R^2 T \propto R^{3/2}$, i.e., the flux {\em increases} with increasing source size.  Thus, for any source in which we may spectrally exclude a photosphere with apparent size $R=\sqrt{27}GM/c^2$, we will more easily be able to do so for immediately larger apparent sizes.
This does not continue indefinitely, however, because the frequency of the spectral peak also decreases with growing photosphere radii.  Eventually $\nu_{\rm peak}$ will fall below the optical range, placing the optical window in the Wien regime, and exponentially suppressing the ability of optical spectra to constrain the photosphere emission.  This occurs when
$T\lesssim 0.9\times10^4~\K$ for an optical spectral constraint at $\nu=5\times10^{14}~{\rm Hz}$, and therefore happens when the apparent photosphere size grows above\footnote{Despite being exponentially suppressed, the flux above the peak does not immediately drop to zero, and thus it is not actually sufficient for the peak to drop below the optical.  Properly addressing $B_\nu$ within the Wien regime function further expands the photosphere by roughly an order of magnitude, and thus our estimate of the $R$ for which the photosphere emission could evade the optical spectral constraints is extremely conservative in practice.}
\begin{equation}
    R \gtrsim \sqrt{27} \frac{GM}{c^2} \left(\frac{T}{10^{4}~\K}\right)^{-2}
    \approx 520 \frac{GM}{c^2} \left(\frac{T}{10^{5}~\K}\right)^{-2}.
\end{equation}
Hence, evading the optical spectral constraints requires apparent source sizes 10--1000 times that of the comparable black hole, which may almost certainly already be excluded via the impact such a photosphere would have on the hard X-ray emission from the accretion flow itself.

\subsection{Photosphere Spectra Uncertainty Estimation}
\label{sec:4-unc}

\begin{figure*}[ht]
    \centering
    \includegraphics[width=\textwidth]{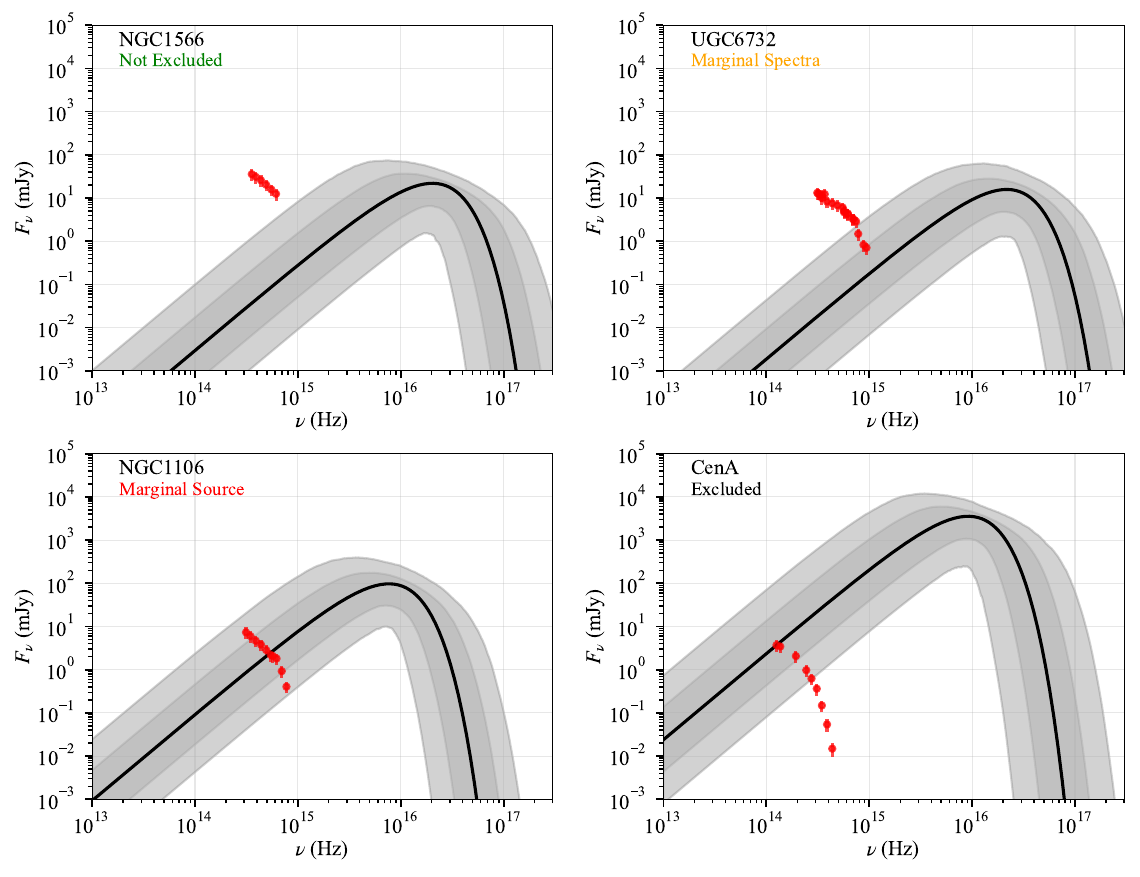}
    \caption{The gray regions are the 68th and 95th percentiles. The four plots show one source as an example from each category. The red points are the binned spectra with error bars. The black curved peak is the expected thermal emission for each source. The excluded source's spectra is significantly below the expected thermal bump and the lower 2$\sigma$ band. The marginal source and marginal spectra are both within the 2$\sigma$ regions. The only source that was classified as not excluded is above the expected the upper 2$\sigma$ band but has limited coverage.
    }
    \label{fig:SED}
\end{figure*}

Many of the source properties required to estimate the putative accretion-powered photosphere spectrum have significant associated uncertainties.  
To estimate the associated uncertainty in the derived photosphere spectra, we perform Monte Carlo error propagation for each source in our sample. 
5000 sample spectra are produced assuming each input parameter listed in \autoref{sec:3-APPE} are distributed normally with the errors described in \autoref{sec:3}.
The resulting photosphere spectra and associated uncertainties in \autoref{fig:SED} for four illustrative examples, where the dark and light gray shaded regions represent the $1\sigma$ ($68\%$) and $2\sigma$ ($95\%$) confidence intervals.

\subsection{Spectral Comparison \& Evaluation Criteria}
\label{sec:4-crit}

For each AGN in \autoref{table1}, the expected photosphere spectra, with associated uncertainties, are directly compared with the BASS reported optical spectra as described in \autoref{sec:3-spec}.  Based on this comparison, each source was assigned to one of the five following categories:

\begin{description}
\item[Detected] Sources with rising optical spectra that qualitatively match the predicted photosphere spectrum prediction within the 95\% region.

\item[Excluded] Sources with optical spectra that fall below the predicted thermal emission and below the lower bound of the 95\% interval were classified as excluded. A minimum of three spectral points with associated error bars needed to be below the lower bound to qualify. These objects conclusively rule out the existence of the photosphere emission.

\item[Marginal Source] Sources with optical spectra that fall within the 95\% region, with multiple spectral points below the median photosphere spectrum prediction, but have fewer than three spectral points below the lower bound of the 95\% interval.
For these objects, moderately better constraints on the source parameters can conclusively exclude photosphere emission.

\item[Marginal Spectra] Sources with optical spectra with at least one point that falls below the upper 95\% region, but not below the median photosphere spectrum prediction.  For these objects, additional optical spectra, which often exhibit significant variability among observation epochs, may conclusively exclude photosphere emission.

\item[Not Excluded] Sources with optical spectra that lay entirely above the upper bound of the 95\% region.  In these cases the predicted photosphere spectrum lies well below that observed, and may be hidden underneath the emission from the accretion flow.
\end{description}

Formally, the Marginal Spectra/Source categories are subsets of the Not Excluded category, but we highlight them to indicate which sources might most benefit from future study.  

By virtue of the photosphere spectra universally appearing in the Rayleigh-Jeans regime at optical wavelengths, we need not engage in detailed spectral fitting.  Rather, it is often sufficient to compare the flux magnitudes and spectral indexes; while potentially more sensitive, more complicated analyses also run the risk of obscuring the practical results.

\subsection{Summary of Comparison Results}
\label{sec:4-comp}

We do not find any examples of an optical spectrum consistent, even qualitatively, with the accretion-powered photosphere predictions among 104 sources in our subsample of the BASS AGN catalog.  This is strongly supported by the fact that we find no evidence for inverted optical spectra ($\alpha>0$) within our sample.  Thus, we now turn to the question of where we can conclusively exclude such a feature.

\autoref{fig:SED} shows a single illustrative source from each of the first four (non-detection) categories listed in \autoref{sec:4-crit}.  The associated classification for each source is listed in \autoref{table1}.  

Of the 104 sources, 39 fall into the excluded category, and thus an accretion-powered photosphere is conclusively absent in these objects.  Much smaller black hole masses, lower inferred accretion rates, or significantly larger distances than implied by the BASS uncertainties could impact this conclusion individual sources.

Of the remaining 65 objects, only one falls into the not excluded category. The implication is that for the vast majority of objects in our BASS subsample that cannot be excluded currently, modest improvements in the measurement of the AGN properties and/or optical spectra from additional observation epochs may be sufficient to formally exclude photosphere emission.  (We discuss increasing the spectral coverage in the following section.)

\begin{figure}[ht]
 \label{Fig:Angular}
    \centering
    \includegraphics[width=\columnwidth]{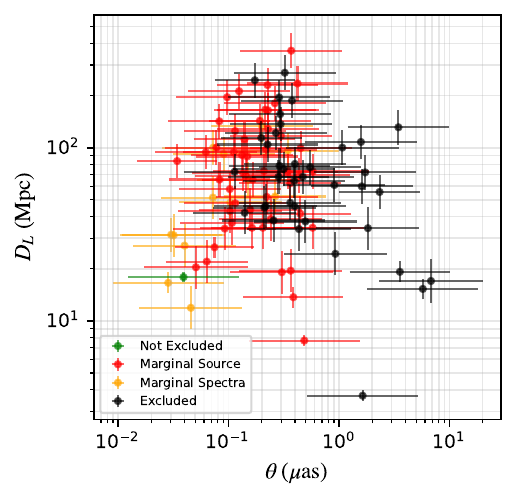}
    \caption{Angular size versus luminosity distance. As the angular size of an AGN increases, it becomes more likely to be classified as Excluded. In contrast, luminosity distance does not appear to show a clear relationship with the classification. The source parameter comparison plots color coded based on their classification: Excluded (black), Marginal Source (red), Marginal Spectra (orange), Not Excluded (green).}
\end{figure}

We further explored the dependence of the resulting categories on several of the source parameters listed in Table \ref{table1} to explore the potential trends, contextualize the nature of the constraints, and guide future observations.  Despite varying substantial across mass, distance, and luminosity, we found a significant trend on a single parameter: the angular apparent source size $\theta$.  We show the distribution of sources in the $\theta$-$D_L$ parameter space in \autoref{Fig:Angular}, with objects color coded according to category.  

Accretion-powered photosphere with larger angular sizes are more likely to be excluded; in no source for which $\theta>1~\muas$ is such a photosphere still possible.  This conclusion is independent of source luminosity or distance.  The sensitivity to source size is a natural consequence of \autoref{eq:Fnu2}, the narrow distribution of photosphere temperatures, and the modest range of optical fluxes in our BASS subsample.  The lack of dependence on source distance may arise from survey flux limits in both the optical and X-ray.

\subsection{Implications for Future Observations}
\label{sec:4-future}

\begin{figure*}[htb!]
    \centering
    \includegraphics[width=\textwidth]{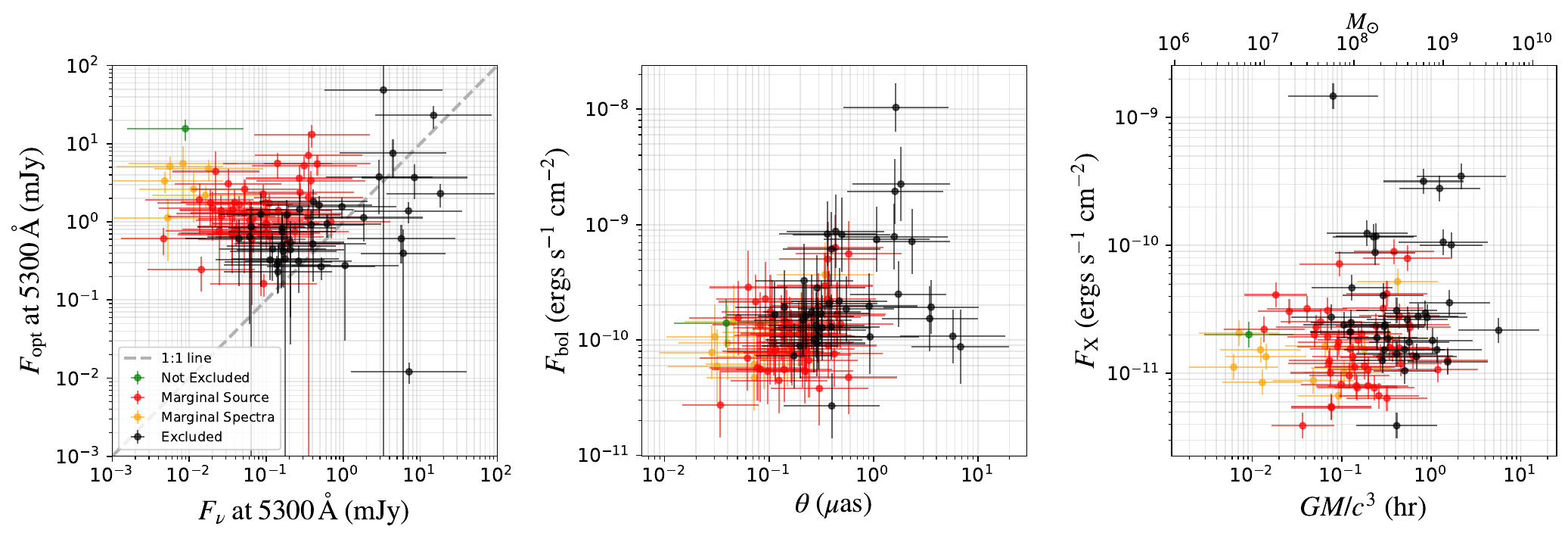}
    \caption{The source parameter comparison plots color coded based on their classification: Excluded (black), Marginal Source (red), Marginal Spectra (orange), Not Excluded (green). Left: The predicted thermal flux versus the observed optical flux density at 5300\,\AA. Middle: The angular size $\theta$ of the thermal region versus bolometric flux. Right: The observed X-ray flux as a function of characteristic timescale and corresponding mass in the upper axis. The error bars represent the source parameter uncertainties and systematic uncertainties mentioned in \autoref{sec:3}.
    }
    \label{fig:comparison_plots}
\end{figure*}

For many of the sources in our subsample, the existence of an accretion-powered photosphere may be conclusively addressed with modest improvements in the measured source properties.  Here we discuss what the general trends in the marginal source characterizations, illustrated in \autoref{fig:comparison_plots}, might imply for three specific classes of potential future observations.

\subsubsection{Additional Spectral Measurements}

As discussed in \autoref{sec:3-spec}, the optical spectra of the AGNs in our subsample of the BASS catalog are strongly variable.  This suggests that simply re-observing many of these sources may exclude accretion-powered photospheres.
The leftmost panel of \autoref{fig:comparison_plots} shows the observed $5300~\mAA$ flux in comparison to that predicted from an accretion-powered photosphere for the sources in our subsample.  As anticipated, sources with high predicted photosphere fluxes and low observed optical fluxes (lower right corner) are predominantly ruled out. However, it is also clear that many more sources have optical fluxes are within less than 1~dex of being excluded.  Therefore, additional spectral observations have a significant probability of finding optical fluxes that are sufficiently low to probe for accretion-powered photospheres.

Because the putative photosphere spectra typically peaks in the EUV, observational constraints on them at UV wavelengths could be far stronger than those we report here.  This expectation arises from two spectral scalings:
First, the spectrum from an accretion-power photosphere increases in the Rayleigh-Jeans regime as $\nu^2$, and thus will be 2-3 orders of magnitude brighter near the peak (see, e.g., \autoref{fig:SED}). Second, all of the sources in our subsample exhibit a optical spectra that decline rapidly with frequency. For a typical spectral index of $\alpha\sim-2.4$, the EUV spectral intensity would be 3 orders of magnitude dimmer at the photosphere peak frequency. Thus, for $\lambda\gg300~\mAA$, as wavelength decreases our test benefits as $\propto \lambda^{\alpha-2}\sim\lambda^{-4.4}$ from simultaneously increasing photosphere signal and decreasing foreground accretion-flow emission.  
Existing and planned UV capabilities can already leverage this wavelength scaling.  
For example, the Cosmic Origins Spectrograph (COS) on the Hubble Space Telescope (HST) can access wavelengths as short as $900~\mAA$, at which we would expect the flux ratio be larger by $\sim 3$ orders of magnitude.  Future UV missions (e.g., CASTOR and UVEX) would provide similar enhancements.  This would be sufficient to conclusively detect or exclude accretion-powered photospheres in all sources listed in \autoref{table1}.

\subsubsection{Direct Size Constraints}

A key assumption in our analysis is that the photosphere size is sufficiently small.  As we argue at the beginning of \autoref{sec:4}, for the vast majority of our sample, this requires only that $R\lesssim10^3 M$.  Nevertheless, with the advent of horizon-resolving imaging enabled by millimeter-wavelength very-long baseline interferometry (mm-VLBI), it is now possible to consider directly constraining the photosphere size.

In the middle panel of \autoref{fig:comparison_plots}, we show the assumed putative photosphere angular size ($2\sqrt{27} GM/c^2D_A=10.4 GM/c^2D_A$) along with the bolometric flux.
Photospheres have been excluded for all objects larger than $1~\muas$.  The largest of these present an angular size just smaller than $10~\muas$, and marginally resolvable by the Event Horizon Telescope (EHT) and future instruments.  However, all objects in our sample would be easily resolvable by ground-based mm-VLBI if their photosphere were large enough to evade the optical flux constraints.  The ability to do so is limited, in that case, by the brightness temperature at mm wavelengths.

We have not surveyed the literature for mm-wavelength observations of sources in our subsample.  Naively extrapolating the optical spectra to 1~mm using the optical spectral indexes produces extremely large mm fluxes that are certainly not realized, implying the existence an intervening spectral break.  Assuming a flat spectrum below the spectral break and a constant spectral index above, the measured $\Fbol$ in combination with $\Fopt$ and $\alpha$ yields a more credible estimate of the flux at mm-wavelengths (see \autoref{app:flux}),
which typical ranges between $0.01$-$1~{\rm Jy}$ for our sources.  For 50 of the 104 objects in our sample, 22~GHz fluxes are published in \citet{Magno2025}, and range from 0.1~mJy to 1~Jy, suggesting that the SED isn't flat at 1.4~cm.  Hence, while highly uncertain, for many objects the mm flux could be sufficiently large to detect these sources with EHT and well within the capacity of other very-long baseline arrays (e.g., the GMVA, VLBA, EVLN, KaVA, LBA).

\begin{deluxetable*}{llccccr}
\tablecaption{Reported VLBI upper limits on the apparent photosphere sizes.\label{tab:vlbi_sizes}}
\tabletypesize{\scriptsize}
\tablehead{
\colhead{BASS} & \colhead{Common} & \colhead{} & \colhead{$\nu$} & \colhead{Flux} & \colhead{Size Limit} & \colhead{}  \\[-1em]
\colhead{ID} & \colhead{Name} & \colhead{Instr.} & \colhead{(GHz)} & \colhead{(mJy)} & \colhead{(mas)} & \colhead{Ref.}
}
\startdata
140 & NGC 1052 & GMVA & 22 & $565 \pm 85$ & $0.35$ & \citet{Baczko2016} \\
173 & NGC 1275 & GMVA & 86 & $(1.0\pm0.2)\times10^3$ & 0.04 & \citet{Paraschos2022}\\
239 & UGC 03157 & EVN & 4.8 & $>3$ & 0.25 & \citet{Parra2010}\\
308 & NGC 2110 & VLBA & 8.4 & $29.5 \pm 1.5$ & 0.72 & \citet{Mundell2000}\\
325 & Mrk 3 & VLBA & 1.6 & $9.4\pm0.9$ & 38 & \citet{Kukula1999}\\ 
436 & NGC 2655 & VLBA & 5 & $0.80\pm0.08$ & 1.8 & \citet{Cheng2025}\\ 
548 & NGC 3718 & EVN & 1.67 & $4.7 \pm 0.3$ & 7 & \citet{Krips2007} \\
579 & NGC 3998 & KaVa & 22 & $ 127 \pm 19$ & 0.99 & \citet{Baek2019} \\
593 & NGC 4138 & EVN & 5.0 &  0.75 & 1.3 & \citet{Bontempi2012} \\
609 & NGC 4258 & VLBA & 22 & $15 \pm 2 $ & 12 & \citet{Cecil2000} \\
665 & NGC 5033 & EVN & 5 & $0.76 \pm 0.08$ & 3.5 & \citet{Giroletti2009} \\
671 & Cen A & EHT & 228 & $\sim2\times10^3$ & 0.020 & \citet{Janssen2021}\\
682 & NGC 5252 & EVN & 1.7 & $3.6 \pm 0.2$ & 3.9 & \cite{Yang2017} \\
841 & NGC 6240 & EVN & 5 & 2.85  & 5.2 & \citet{Hagiwar2011} \\
841 & NGC 6240N & EVN & 5 & 5.79 & 3.4 & \citet{Hagiwar2011} \\
1142 & NGC 7213 & LBA & 8.425 & $57.6 \pm 1.3 $ & 2.8 & \citet{Blank2005} \\
1184 & NGC 7479 & VLBA & 5 & $285.3 \pm 22.0$ & 1.6 & \citet{Laine2026} \\
\enddata
\end{deluxetable*}

Indeed, we note that many sources within our subsample have already been imaged by VLBI arrays.  For example, EHT has imaged Centaurus A \citep{Janssen2021}.  Our assumed photosphere size would subtend an angle of $1.6~\muas$, well below the size of features resolved by EHT.  However, this is not the relevant scale in these sources for the reasons described previously.  That is, to avoid the optical flux limits altogether would require the photosphere to grow in size by 3 orders of magnitude, placing it on mas scales.  Such a large central object is directly amenable by existing instruments, and has already been excluded by EHT \citep{Janssen2021}, but also VLBA \citep{Tingay2001, Tingay1998}, VSOP \citep{Horiuchi2006}, LBA \citep{Ojha2010}, SPT-APEX \citep{Kim2018}.
A number of mas-scale size constraints for the sources listed in \autoref{table1} exist within the literature, which we collect in \autoref{tab:vlbi_sizes}.

\subsubsection{Indirect Size Constraints}
\label{subsec:Indirect_Size_Constraints}
The timescales of source variability places an independent constraint on the size of the emission region \citep{Celotti1998,Morgan2010,Blackburne2011}.  Insofar as the emission arises near the central object, it may be used in principle to constrain the size of a putative photosphere.
This is complicated by the assumed source of the variability, the relationship between global and microscopic scales (e.g., the turbulent spectrum), and what sets the variability timescales on each scale (e.g., orbital motion, light crossing time, etc.).  Nevertheless, we present an initial attempt to identify features in the power spectrum of archival X-ray light curves for Cen A in \autoref{App:Size_estimate}. 

The timescale relevant for constraining the size of the photosphere is the light crossing time of central object.  This is a factor of order $10 GM/c^3$, depending on the details of the central object (e.g., the period as measured at infinity of a half orbit around the photon sphere of a Schwarzschild black hole is $\pi \sqrt{27}GM/c^3$).  Because the X-ray emission is expected to arise, in part, from the innermost portions of the accretion flow, the $S/N$ with which light curve variability may be constrained is $\propto F_X^{1/2}$.  Therefore, in the rightmost panel of \autoref{fig:comparison_plots} we show the typical timescale in comparison to the X-ray flux from the BASS catalog for our subsample.  Those objects that fall into our excluded category typically have variability timescales of hours, and are systematically brighter than the others.

\section{Conclusions}
\label{sec:C}

We report a search for accretion powered photospheres in a large sample of supermassive black hole candidates.  While black holes possess an event horizon that prevents the formation of such a feature, photospheric emission is a generic feature for many black hole alternatives, including boson stars, gravastars, fermion stars, and naked singularities.  Thus, we are ultimately placing constraints on the viability of a wide class of black hole mimickers.

We select a subsample of 104 low-luminosity AGNs from the BASS catalog, corresponding to the radiatively inefficient and optically thin sources for which the photosphere emission would be observable should it exist.  The BASS catalog is a unique resource in that it combines estimated source properties necessary to predict the photosphere spectral signature (mass, distance, bolometric flux) and the optical spectra that constrain its existence.

We find no detections of an accretion powered photosphere, i.e., no spectra with inverted spectral index and similar amplitude to that predicted by our accretion powered photosphere model.  This is independent of the large uncertainties in source properties.

In 39 of 104 sources, the optical flux measurements fall well below the predicted photosphere predictions, and therefore we can conclusively exclude the presence of such a spectral feature.  Therefore, in these sources, we provide strong evidence for horizons.

For the remainder of our sample, we cannot conclusively exclude the existence of a thermal photosphere component, i.e., the optical flux limits lie above the predicted photosphere predictions.  However, in all but 1 of the remaining 65 objects, a modest decrease in the optical spectrum (due to accretion flow variability), improved source property constraints, or higher-frequency spectra (UV) would prove decisive.  

A key assumption in our analysis is the intrinsic photosphere size: larger photospheres are cooler and their spectral signature could more easily be hidden under the direct emission from the surrounding accretion flow.  However, to evade the current optical flux constraints, the photosphere would need to grow to sizes of $\sim10^4 GM/c^2$, large enough to subtend mas angular scales, amenable to direct imaging.  For 17 of the 104 objects in our sample, archival VLBI observations can already rule out such large source sizes; for many of the remainder, radio flux estimates suggest that direct imaging may be feasible. 
Such large photospheres would also directly impact the surrounding accretion flow, substantially modifying the hard X-ray emitting regions responsible for the BASS X-ray luminosity estimates.  Thus, source size may also be constrained indirectly by light curve variability and spectral modeling of the accretion flow.

We have not made any attempt to extend our photosphere predictions to the spinning spacetimes.  While this may modify the quantitative predictions of the surface spectrum, we do not believe that it will substantially alter our results.  Nevertheless, incorporating spin significantly complicates the dynamics of accreting gas and enhance the class of geodesics that can be explored, and we leave this for future work.

Ways to evade the implications of our constraint for the fundamental physics of black holes exist.  Black hole mimickers that absorb matter and energy sufficiently rapidly, e.g., non-local interactions that can remove the kinetic and internal energy of the accreting gas, will necessarily not produce the requisite photosphere structure.  Additionally, models in which an astronomically long-lived apparent horizon develops could hide the photosphere sufficiently to prevent its detection today.  However, such limitations are already represent extreme constraints on black hole alternatives, posing significant challenges for physically motivated models.

\begin{acknowledgments}
This work was motivated by a presentation by Richard Mushotzky at ``McNamara@65: Understanding Feedback in Galaxies and Clusters'' held at the University of Waterloo from May 27-30, 2024 in celebration of the life and work of Brian McNamara.  We thank both Richard and Brian for their inspiration.
The Perimeter Institute for Theoretical Physics partially supported this work. Funding for research at the institute is provided by the Department of Innovation, Science and Economic Development Canada, and the Ministry of Economic Development, Job Creation and Trade of Ontario, both of which are branches of the Government of Canada.  Additionally, A.E.B. receives further financial support for this research through a Discovery Grant from the Natural Sciences and Engineering Research Council of Canada.
\end{acknowledgments}

\newpage
\appendix
\section{One-zone optical depth estimates}
\label{app:tau}

We estimate the source optical depth to synchrotron absorption at EUV wavelengths via a simple equipartition model.  We assume the presence of centrally concetrated population of nonthermal electrons with a power-law energy distribution and a density that varies as a radial power law, appropriate for RIAFs or Bondi flows,
\begin{equation}
    \frac{dn}{d\gamma} = C \gamma^{-p} r^{-a}.
\end{equation}
The magnetic field is assumed to be in equipartion with the energy density of the nonthermal electrons, i.e., 
\begin{equation}
\begin{aligned}
    u_B &= \beta^{-1} u = \beta^{-1} \int_{\gamma_{\rm min}}^{\gamma_{\rm max}} \gamma m_e c^2 \frac{dn}{d\gamma} d\gamma\\
    &\sim
    C m_e c^2 \beta^{-1} \gamma_{\rm min}^{2-p} r^{-a}.
\end{aligned}
\end{equation}
where $\beta$ is analogous to the standard plasma $\beta$ and we set $\beta\approx10^2$.  Therefore,
\begin{equation}
    \nu_B \equiv \frac{eB}{2\pi m_e c} 
    = \left(\frac{2\gamma_{\rm min}}{\beta}\right)^{1/2} \nu_P,
\end{equation}
where 
\begin{equation}
    \nu_P^2 \equiv \frac{4\pi e^2}{m_e} C \gamma_{\rm min}^{1-p} r^{-a} 
\end{equation}
is the square of the plasma frequency associated with the nonthermal component.

The coefficient $C$, and therefore $\nu_P$ at $r_{\rm in}$, is set by requiring that the observed flux in the EUV arises from this population of electrons.  For this we note that the emissivity and absorption coefficient for synchrotron emission from a power law electron distribution like that we assume here are approximately given by,
\begin{equation}
\begin{aligned}
    j_\nu &\sim \frac{\pi}{4 \sqrt{3}}  \frac{m_e c^2}{\lambda^3} \left(\frac{\nu_P}{\nu}\right)^2 \left(\frac{\nu_B}{\nu}\right)^{(p+1)/2} \Gamma_j \\
    \alpha_\nu &\sim \frac{\pi}{4\sqrt{3}} \frac{1}{\lambda} \left(\frac{\nu_P}{\nu}\right)^2\left(\frac{\nu_B}{\nu}\right)^{(p+2)/2} \Gamma_\alpha 
\end{aligned}
\end{equation}
where $\Gamma_j$ and $\Gamma_\alpha$ are dimensionless functions of $p$ that are of order unity for all $p$ that are relevant here.  Upon inserting the equipartition value for $\nu_B$, the emissivities becomes
\begin{equation}
    j_\nu \sim \frac{m_e c^2}{\lambda^3} \left(\frac{2\gamma_{\rm min}}{\beta}\right)^{(p+1)/4} \left(\frac{\nu_P}{\nu}\right)^{(p+5)/2},
\end{equation}
and the flux due to synchrotron emission from the accretion flow is
\begin{equation}
\begin{aligned}
    F_\nu 
    &= \frac{1}{D^2} \int_{r_{\rm in}}^\infty dr\, 4\pi r^2 j_\nu\\
    &\sim \frac{4\pi r_{\rm in}^3}{D^2}
    \frac{m_e c^2}{\lambda^3} \left(\frac{2\gamma_{\rm min}}{\beta}\right)^{(p+1)/4} \left(\frac{\nu_{P,\rm in}}{\nu}\right)^{(p+5)/2},
\end{aligned}
\end{equation}
where $\nu_{P,\rm in}$ is the plasma frequency at $r_{\rm in}$.  Thus,
\begin{equation}
    \frac{\nu_{P,\rm in}}{\nu}
    \sim 
    \left[ \frac{D^2}{4\pi r_{\rm in}^3} \frac{F_\nu \lambda^3}{m_e c^2} 
    \left(\frac{\beta}{2\gamma_{\rm min}}\right)^{(p+1)/4} \right]^{2/(p+5)}.
\end{equation}
Using this to set $C$ in terms of $F_\nu$, the optical depth of the innermost region of the accretion flow is
\begin{equation}
\begin{aligned}
    \tau_\nu 
    &= \int_{r_{\rm in}}^\infty dr \, \alpha_\nu\\   
    &\sim \frac{r_{\rm in}}{\lambda} \left(
        \frac{\lambda^3}{r_{\rm in}^3} \frac{D^2 F_\nu}{4\pi m_e c^2}
    \right)^{(p+6)/(p+5)} \left(\frac{2\gamma_{\rm min}}{\beta} \right)^{1/(p+5)}.
\end{aligned}
\end{equation}
Typical spectral indexes are $\sim1.25$, and therefore, $p\sim3.5$.  Therefore, the optical depth is essentially independent of $\gamma_{\rm min}$ and $\beta$ and 
\begin{equation}
    \tau_\nu \sim \frac{\lambda^2 D^2 F_\nu}{4\pi r_{\rm in}^2 m_e c^2}.
\end{equation}
Inserting typical values for $\lambda$, $D$, $F_\nu$ and $r_{\rm in}=2 GM/c^2$, this evaluates at $\nu=10^{15}\,{\rm Hz}$ to
\begin{equation}
\begin{aligned}
    \tau_\nu &\sim 10^{-4} 
    \left(\frac{\nu}{10^{15}\,{\rm Hz}}\right)^{-2} 
    \left(\frac{D}{10^2\,\Mpc}\right)^2\\
    &\qquad\qquad\qquad\quad\times
    \left(\frac{M}{10^8\,M_\odot}\right)^{-2} 
    \left(\frac{F_\nu}{1\,{\rm mJy}}\right).
\end{aligned}
\end{equation}
We evaluate this individually for each source and find that it is generally small.

\section{Size estimation using Variability}
\label{App:Size_estimate}

Constraining the size of the emission region is crucial for distinguishing between the presence of an event horizon and its absence, as in the case of an accretion powered photosphere. One way to constrain the size of the emitting region is by analyzing the short-timescale variability. In principle, the break frequency, $t\propto {\nu_{b}^{-1}}$, in a periodogram reveals the size of the compact region through the relation $R \leq ct$, where $t$ is the light-crossing time across the emitting region. This allows us to place an upper limit on the physical size of the emitting region. Within our sample, Centaurus A (Cen A) is an ideal candidate for this analysis given its proximity and brightness.

We analyzed archival X-ray data spanning $0.8 \text{--} 60~\text{keV}$, utilizing observations from HEAO-1, EXOSAT, and EINSTEIN (HEAO-2) retrieved from the XAMIN Web Interface. In addition, we analyzed raw Chandra HETG\footnote{HETG stands for High Energy Transmission Grating: this instrument is ideal for very bright sources to mitigate photon pile up} observations from the HEASARC archive, covering the $0.4 \text{--} 10~\text{keV}$ range.

\begin{figure}[ht]
    \centering
    \includegraphics[width=\columnwidth]{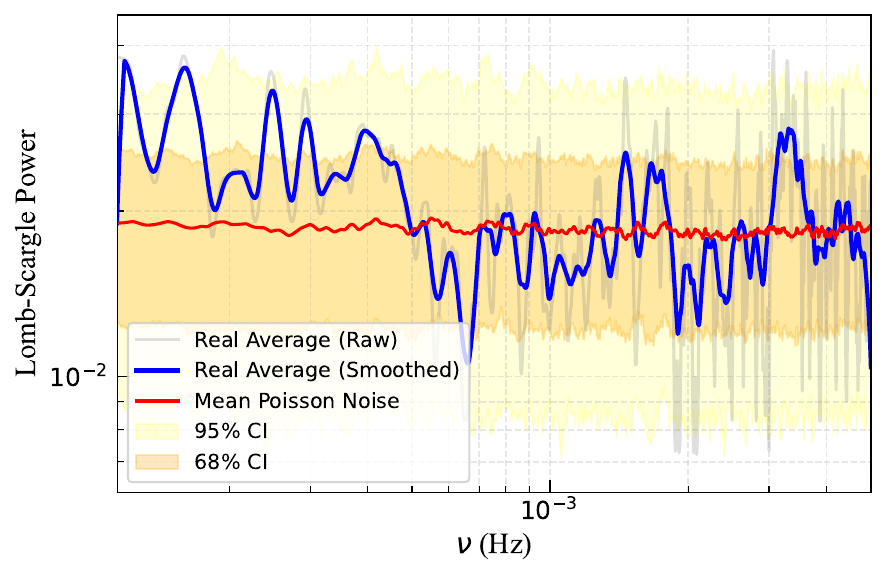}
    \includegraphics[width=\columnwidth]{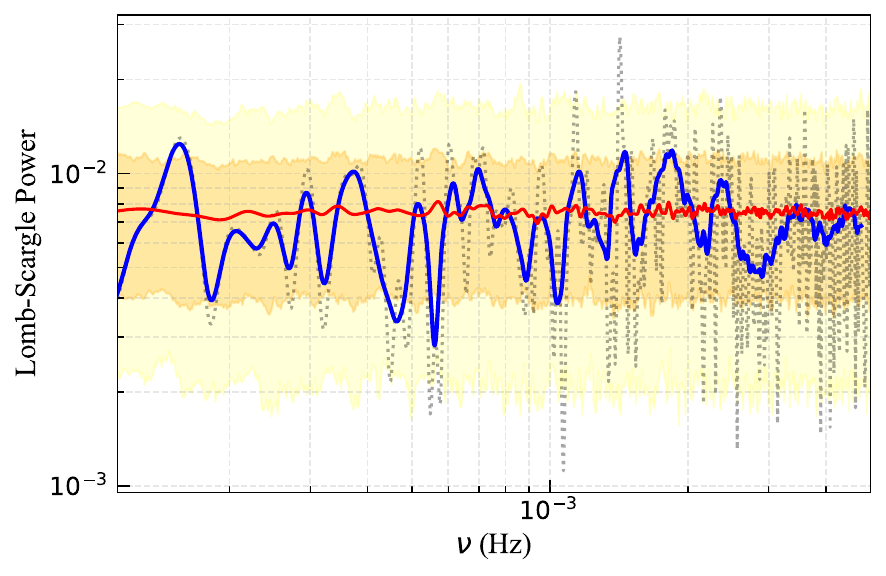}
    \caption{Top: Averaged periodogram of archival X-ray data from the HEAO-1, EXOSAT, and EINSTEIN missions. Bottom: Averaged periodogram of dispersed X-ray data from the Chandra HETG observations. The blue signal represents the mean intrinsic source variability, while the red signal shows the mean simulated Poisson noise. The orange and yellow intervals represent the 68th and 95th percentiles, respectively.}
    \label{Fig: XAMIN&Chandra}
\end{figure}

The analysis consisted of converting the light curves with $100~\text{s}$ bins from each observation into Lomb-Scargle periodograms, then taking the average. The upper plot in Fig.~\ref{Fig: XAMIN&Chandra} shows the averaged periodogram derived merely from HEAO-1, EXOSAT, and EINSTEIN data. To distinguish the intrinsic source variability from the white noise floor, we then performed a Poisson noise simulation with $2 \sigma$ confidence intervals. The lower plot shows the averaged periodogram from Chandra observations only. Since Chandra did not provide a pre-processed, cleaned and background subtracted light curves, the raw event files (evt2) were processed using the CIAO-4.17 software and visualized in DS9. To avoid detector saturation, we extracted the light curves specifically from the dispersed grating arms (HEG\&MEG) and applied background subtraction in DS9. Once the light curve files were created, the same procedure was followed.

Across both datasets, no break frequency was detected above the white noise floor. Although, the archival data from XAMIN showed a potential break near ~$10^{-3} ~\text{Hz}$, it remained statistically insignificant as it was indistinguishable from the white noise floor. Similarly, the Chandra data variability was comparable to purely white noise. The lack of red noise could be attributed to the need for a broader or harder X-ray band. Since the higher energy X-rays typically originate closer to the central engine, whereas the softer X-rays are more susceptible to contamination from the host galaxy's spectral emission. 

Future projects can look further into the hard X-ray bandwidth ($\geq 10$ keV), since AGNs show stronger variability on shorter timescales at  higher energies. It is mentioned in \citet{Tortosa_2023} that there is an anti-correlation between black hole mass and variability amplitude. More massive and luminous black holes are less likely to exhibit strong variability as a direct consequence of longer light crossing times across larger emitting regions. This correlation helps explain why strong variability was undetected on short timescales in Cen A given that it's a moderately large black hole ($M_{BH} \sim 5.8\times10^{7}\, M_\odot$).

\section{Millimeter-wavelength Flux Estimates}
\label{app:flux}

We make a simple gross estimate of the millimeter flux density motivated by the BLANDFORD-KONIGLE (197?) model of self-absorbed jet sources.  We assume that the spectral energy distribution (SED) is given by
\begin{equation}
    F_\nu = F_b \begin{cases}
        1 & \nu<\nu_b\\
        (\nu/\nu_b)^\alpha & \nu\ge\nu_b,
    \end{cases}
\end{equation}
where $\nu_b$ is the location of the spectral break, below which the spectrum is flat.  We will further assume that $\nu_b\ll\nu_{\rm opt}$, where $\nu_{\rm opt}=5.7\times10^{14}~{\rm Hz}$, and $\alpha<-1$ as found for our sample (see \autoref{sec:3-spec}).  For the purposes of our estimate, we will ignore modifications to the SED above optical wavelengths (e.g., inverse-Compton or bremsstrahlung components) which may contribute substantially to $\Fbol$ in practice, and this must be born in mind.

Observational inputs include the flux at $\nu_{\rm opt}$, $\Fopt$, the optical spectral index $\alpha$, and the total bolometric flux estimate, $\Fbol$.  These inputs are related to our assumed SED via
\begin{equation}
    \Fopt = F_b (\nu_{\rm opt}/\nu_b)^\alpha
\end{equation}
and
\begin{equation}
    \Fbol = \int_0^\infty F_\nu \, d\nu
    = \frac{\alpha F_b \nu_b}{(1+\alpha)},
\end{equation}
which may be solved to obtain expressions for the location of the spectral break and the low-frequency flux:
\begin{equation}
    \nu_b = \nu_{\rm opt} \left[ 
        \frac{\alpha \Fopt \nu_{\rm opt}}{(1+\alpha) \Fbol}
    \right]^{-1/(1+\alpha)},
\end{equation}
and
\begin{equation}
    F_b
    = 
    \Fopt^{1/(1+\alpha)} \left[\frac{(1+\alpha) \Fbol}{\alpha \nu_{\rm opt}} \right]^{\alpha/(1+\alpha)}.
\end{equation}
For all $\nu<\nu_b$, this $F_b$ provides the desired estimate of the spectral flux density.

\bibliographystyle{aasjournalv7.1}
\bibliography{references}

\end{document}